\documentclass[12pt]{article}

\usepackage[truedimen,margin=34mm]{geometry}

\usepackage{mathrsfs}
\usepackage{amssymb}
\usepackage{amsmath}
\usepackage{amsthm}
\usepackage[pdftex]{graphicx}
\usepackage[pdftex]{color}
\usepackage{booktabs}
\usepackage{setspace}
\usepackage{natbib}
\usepackage{multirow,color}

\usepackage{titlesec}
\titleformat*{\section}{\large\bfseries}
\titleformat*{\subsection}{\it}

\newtheorem{theorem}{Theorem}

\theoremstyle{definition}

\newtheorem{remark}{Remark}

\newcommand{\bx}{{\mbox{\boldmath$x$}}}
\newcommand{\bX}{{\mbox{\boldmath$X$}}}
\newcommand{\bA}{{\mbox{\boldmath$A$}}}
\newcommand{\bV}{{\mbox{\boldmath$V$}}}

\newcommand{\bh}{{\mbox{\boldmath$h$}}}

\newcommand{\bB}{\mathbf{B}}
\newcommand{\bC}{\mathbf{C}}

\newcommand{\bD}{\mathbf{D}}

\newcommand{\la}{\lambda}
\newcommand{\ep}{\varepsilon}

\newcommand{\bbeta}{{\mbox{\boldmath$\beta$}}}
\newcommand{\bbe}{{\mbox{\boldmath$\beta$}}}
\newcommand{\bbeh}{\hat{\bbeta}}

\newcommand{\bVh}{\hat{\bV}}
\newcommand{\Vh}{\hat{V}}

\newcommand{\beh}{\hat{\beta}}

\newcommand{\E}{{\rm E}}
\newcommand{\Var}{{\rm Var}}
\newcommand{\barY}{\bar{Y}}
\newcommand{\hbarY}{\hat{\barY}_{\rm reg}}

\DeclareMathOperator*{\plim}{plim}

\title{\bf An Approximate Bayesian Approach to Model-assisted Survey Estimation with Many Auxiliary Variables  }
\author{
Shonosuke Sugasawa\thanks{Center for Spatial Information Science, The University of Tokyo}
\and
Jae Kwang Kim\thanks{Department of Statistics, Iowa State University}
}
\date{\today}

\begin{document}
%\doublespacing
\maketitle

\thispagestyle{empty}

\bigskip
\begin{abstract}
Model-assisted estimation with complex survey data is an important practical problem in survey sampling. When there are many auxiliary variables, selecting significant variables associated with the study  variable would be necessary to achieve efficient estimation of population parameters of interest. In this paper, we formulate a regularized regression estimator in the framework of Bayesian inference using the penalty function as the shrinkage prior for model selection. The proposed Bayesian approach  enables us to get not only efficient point estimates but also reasonable credible intervals. Results from two limited simulation studies are presented to facilitate  comparison  with existing frequentist methods.
\end{abstract}

\noindent%
{\it Keywords:} Generalized regression estimation; Regularization; Shrinkage prior; Survey Sampling
\vfill
%\nopagenumbering

\newpage

\baselineskip .3in

\setcounter{page}{1}
%\spacingset{1.45} % DON'T change the spacing!

\section{Introduction}

Probability sampling is a scientific tool for obtaining a representative sample from the target population. In order to estimate a finite population total from a target population,  Horvitz-Thompson (HT) estimator obtained from a probability sample satisfies design-consistency and the resulting inference is justified from the randomization perspective \citep{ht1952}.  However, the HT estimator uses the first-order inclusion probability only and does not fully incorporate all available information in the finite population. To improve its efficiency, regression estimation is often used by incorporating  auxiliary information in the finite population. 
  \cite{deville1992}, \cite{fuller2002}, \cite{kim2010}, and \cite{breidt17} present comprehensive overviews of variants of regression estimation in survey sampling.
There are also other directions of improvement on the HT estimator based on  prediction using augmented models  
 \citep[e.g.][]{ZL2003, ZL2005, ZL2015}.

The regression estimation approaches in survey sampling assume a model for the finite population, i.e., the superpopulation model, as
\begin{equation}
y_i =   \bx_i^t \bbeta   + e_i,
\label{1}
\end{equation}
where $y_i$ is a response variable,  $\bx_i$ and $\bbeta$ are vectors of auxiliary variables and regression coefficients, respectively, and $e_i$ is an error term satisfying $\E(e_i)=0$ and $\Var(e_i)=\sigma^2$.
%Model (\ref{1}) is a model for the finite population and is called the  superpopulation model.
The superpopulation model does not necessarily hold in the sample as the sampling design  can be informative \citep[e.g.][]{Pfeffermann1999, little2004}.
%Instead of the linear model (\ref{1}), some extensions to nonlinear models are also considered as in \cite{wu2001}, \cite{breidt05}, and \cite{montanari2005}.
Under the regression superpopulation model in (\ref{1}), \cite{isaki1982} show that the asymptotic variance of the regression estimator achieves the lower bound of \cite{godambe1965}. Thus, the regression estimator is asymptotically efficient in the sense of achieving the minimum anticipated variance under the joint distribution of the sampling design and the superpopulation model in (\ref{1}).

On the other hand, the dimension of the auxiliary variables $\bx_i$ could be large in practice.
Even when the number of observed covariates is not necessarily large, the dimension of $\bx_i$ could be very large once we include polynomial or interaction terms to achieve flexible modeling, as considered in Section \ref{sec:exm}.    
However, in this case, the optimality of the regression estimator is untenable. When  there are many auxiliary variables,   the asymptotic bias of the regression estimator using all the auxiliary variables  is no longer negligible and the resulting inference can be problematic. Simply put, including irrelevant
auxiliary variables can introduce substantial  variability in  point estimation, but its uncertainty is not fully accounted for by the standard linearization variance estimation, resulting in misleading inference.

%the variance estimator   can  underestimate its uncertainty and the resulting inference can be misleading.
%  Furthermore, the resulting regression estimator with many auxiliary variables can lead to extreme weights.

To overcome the problem, variable selection techniques for regression estimation have been considered in literatures \cite[e.g.][]{Silva1997, sarndal2005}.
The classical selection approach is based on a step-wise method. 
However, the step-wise methods will not necessarily produce the best model \citep[e.g.][]{Demp1977} although the potential effect on prediction could be limited.  Another approach is to  employ  regularized estimation of regression coefficients.
% by adding a penalty function to (\ref{OLS}).
For example, \cite{Mc2017}  propose a regularized regression estimation approach based on the LASSO penalty  of \cite{tib1996}.
%Also, \cite{chen2018}  considered the regularized regression estimator using adaptive LASSO of \cite{zou2006}.
%and ridge regularization in (\ref{OLS}).
However, there are two main problems with the regularization approach in regression estimation. 
First, the choice of the regularization parameter is not straightforward under survey sampling when the parameter is strongly related to the selection results.
% The 10-fold cross validation method can be used, but the choice of the subsample under the complex sampling is not totally clear.
 Second, after model selection,  the frequentist inference  is notoriously difficult to make.

In this paper, to overcome the above difficulties, we adopt a Bayesian framework in the regularized regression estimation. 
 We first introduce an approximate Bayesian approach for regression estimation when $p+1= \mbox{dim} (\bx)$ is fixed, using the  approximate Bayesian approach considered in \cite{wang2018}. The proposed Bayesian method fully captures the uncertainty in  parameter estimation for  the regression estimator and has good coverage properties. Second, the proposed Bayesian method is extended to the problem of large $p$ in regularized regression estimation.  By incorporating the penalty function for regularization  into the prior distribution, the uncertainty associated with model selection and parameter estimation is fully captured in the Bayesian machinery.
Furthermore,  the choice of  the penalty parameter can be handled by  using its posterior distribution.
Hence, the proposed method provides a unified approach to Bayesian inference with sparse model-assisted survey estimation.
The proposed method is a calibrated Bayesian \citep{little2012} and it is asymptotically equivalent to the frequentist model-assisted approach for a fixed $p$.

The paper is organized as follows. In Section 2, the basic setup is introduced. In Section 3, the approximate Bayesian inference using   regression estimation is proposed under a  fixed $p$ setup. In Section 4, the proposed method is extended to high dimensional setup by developing  sparse regression estimation using shrinkage prior distributions. In Section 5, the proposed method is extended to non-linear regression models. In Section 6,
results from two limited simulation studies are presented. The proposed method is applied to the real data example in Section 7. Some concluding remarks are made in Section 8.

%----------------------------------------------------------------%
%       Basic setups
%----------------------------------------------------------------%
\section{Basic setup}

Consider a finite population of a known size $N$. Associated with unit $i$ in the finite population, we consider measurement $\{\bx_i, y_i\}$ where $\bx_i$ is the vector of auxiliary variables with dimension $p$ and $y_i$ is the study variable of interest. We are interested in estimating the finite population mean $\barY  = N^{-1} \sum_{i=1}^N y_i$ from a sample selected by a probability sampling design. Let $A$ be the index set of the sample and we observe $\{\bx_i, y_i\}_{i\in A}$ from the sample.
The HT  estimator $\hat{\bar{Y}}_{HT}=N^{-1} \sum_{i \in A} \pi_i^{-1} y_i$, where $\pi_i$ is the first-order inclusion probability of unit $i$, is design unbiased but it is not necessarily efficient.

If the finite population mean $\bar{\bX}= N^{-1} \sum_{i=1}^N \bx_i$ is known, then we can improve the efficiency of $\hat{\bar{Y}}_{\rm HT}$ by using the following   regression  estimator:
$$ 
\hbarY = \frac{1}{N} \sum_{i=1}^N \bx_i^t \hat{\bbeta} 
$$
where   $\bbeh$ is an estimator of $\bbeta$ in (\ref{1}).  
Typically, we use $\bbeh$ obtained by minimizing the weighted quadratic loss
\begin{equation}\label{OLS}
Q(\bbeta) = \sum_{i\in A}\pi_i^{-1}(y_i-\bx_i^t\bbeta)^2,
\end{equation}
motivated from the model (\ref{1}). If an intercept term is included in $\bx_i$ such that $\bx_i^t = (1, \bx_{1i}^t )$, we can express 
%If an intercept term is included in $\bx_i$ such that $\bx_i^t = (1, \bx_{1i}^t )$, then we can express 
\begin{equation} \label{GREG}
\hbarY = \hat{\beta}_0 + \bar{\bX}_1^t  \bbeh_1    =    \hat{N}^{-1}  \sum_{i \in A}  \pi_i^{-1} \left( y_i - \bx_{1i}^t \hat{\bbeta}_1 \right) + \bar{\bX}_1^t  \bbeh_1\end{equation}
where $\hat{N} = \sum_{i \in A} \pi_i^{-1}$ and $\hat{\bbeta}_1$ is given by 
\begin{equation} 
 \hat{\bbeta}_1 = \left\{ \sum_{i \in A} \pi_i^{-1}  (\bx_{1i} - \hat{\bar{\bX}}_{1, \pi}  )^{\otimes 2}  \right\}^{-1} \sum_{i \in A} \pi_i^{-1}  (\bx_{1i} - \hat{\bar{\bX}}_{1, \pi}  ) y_i 
 \label{bhat1}
 \end{equation} 
where 
$ \hat{\bar{\bX}}_{1, \pi}  = \hat{N}^{-1} \sum_{i \in A} \pi_i^{-1} \bx_{1i} $
and $B^{ \otimes 2} = B B'$ for some matrix $B$.

To discuss  some asymptotic properties of $\hbarY$ in (\ref{GREG}), we consider a sequence of finite populations and samples as discussed in Isaki and Fuller (1982), where $N$ increases with $n$. Note that 
\begin{align}
\hbarY - \barY &=   \hat{\barY}_{\pi} - \barY + \Big(\bar{\bX}_1 - \hat{\bar{\bX}}_{1, \pi}  \Big)^t \hat{\bbeta}_1 \notag \notag \\
&=   \hat{\bar{Y}}_{\pi} - \bar{Y} + \left(\bar{\bX}_1 -  \hat{\bar{\bX}}_{1, \pi}     \right)^t{\bbeta}_1 + R_n
\label{eq:expand}
\end{align}
where  $\hat{\bar{Y}}_{\pi} = \hat{N}^{-1} \sum_{i \in A} \pi_i^{-1} y_i$ and 
$$
R_n =  \left(\bar{\bX}_1 -\hat{\bar{\bX}}_{1}  \right)^t\left( \bbeh_1 - {\bbeta}_1 \right)
$$
for any $\bbe_{1}$.
If we choose $\bbe_{1}  = p \lim_{n\to\infty} \hat{\bbeta}_1$ with respect to the sampling probability and  $p=\mbox{dim} (\bx_1)$ is fixed in the asymptotic setup, then we can obtain $R_n = O_p (n^{-1})$ and  safely use the main terms of (\ref{eq:expand}) to describe the asymptotic behavior of $\hbarY$. To emphasize its dependence on $\bbeh_1$ in the regression estimator, we can write $\hbarY= \hbarY ( \bbeh_1 )$.
Roughly speaking, we can obtain
\begin{equation}
 \sqrt{n} \left\{  \hbarY ( \hat{\bbeta}_1 ) - \hbarY ( \bbe_1)  \right\} = O_p (n^{-1/2} p).
 \label{result1}
 \end{equation}
and, if $p=o(n^{1/2})$ then we can safely ignore the effect of estimating $\bbe_1$ in the regression estimator. See Supplementary Material for a sketched  proof of (\ref{result1}).

If, on the other hand, the dimension $p$ is larger than $O(n^{1/2})$, then we cannot ignore the effect of estimating $\bbe_1$. In this case, we can consider using  some variable selection idea to reduce the dimension of $\bX$.
For variable selection, we may employ techniques of regularized estimation of regression coefficients.  The regularization method can be described as finding
\begin{equation}
(\beh_0^{(R)},\hat{\bbeta}_1^{(R)}) = \mbox{argmin}_{\beta_0,\beta_1} \{  Q( \bbeta) + p_\lambda ( \bbeta_1) \},
\label{regular}
\end{equation}
where $Q( \bbeta)$ is defined in (\ref{OLS}) and $p_{\lambda} ( \bbeta_1)$ is a penalty function with  parameter $\lambda$.
%For the Lasso-type regression estimation, \cite{Mc2017} used $p_\lambda = \lambda \sum_{j=1}^p \left| \beta_j \right| $.
Some popular penalty functions are presented in Table 1.
Once the solution to (\ref{regular}) is obtained, then the regularized regression estimator is given by
\begin{equation} \label{rreg}
\hbarY ( \bbeh_1^{(R)} )  = \bar{\bX}_1^t  \bbeh_1^{(R)}  + \frac{1}{\hat{N}} \sum_{i \in A} \frac{1}{\pi_i} \left( y_i - \bx_{1i}^t \hat{\bbeta}_1^{(R)}  \right).
\end{equation}
Statistical  inference with the regularized regression estimator in (\ref{rreg})  is not fully investigated in the literature. For example, \cite{chen2018}  consider the regularized regression estimator using adaptive LASSO of \cite{zou2006}, but they assume that the sampling design is non-informative and the uncertainty in  model selection is not fully incorporated in their inference.
Generally speaking, making inference after model selection under superpopulation frequentist framework is difficult.
The approximated Bayesian method introduced in the next section will capture the full uncertainty in the Bayesian framework.

\begin{table}
		\caption{Popular penalized regression methods }\label{Prop1}
		\centering
		\begin{tabular}{c|c|c}
			\hline
	Method &  Reference & Penalty function   \\
\hline
Ridge & \cite{Hoerl70}  & $p_\lambda( \bbeta) = \lambda \sum_{j=1}^p  \beta_j^2  $   \\
LASSO & \cite{tib1996} & $p_\lambda( \bbeta) = \lambda \sum_{j=1}^p \left| \beta_j \right| $  \\
Adaptive LASSO & \cite{zou2006}  & $p_\lambda( \bbeta) =\lambda \sum_{j=1}^p \left( \left| \beta_j \right| / \left| \hat{\beta} _j \right| \right)  $ \\
%SCAD &  & \\
Elastic Net & \cite{zou2005} &  $p_\lambda( \bbeta) =\lambda_1 \sum_{j=1}^p \left| \beta_j \right|  + \lambda_2 \sum_{j=1}^p \beta_j^2  $\\
			\hline
		\end{tabular}
	\end{table}

%----------------------------------------------------------------%
%       Approximate Bayesian regression
%----------------------------------------------------------------%

\section{Approximate Bayesian survey regression estimation} \label{sec:ABR}

Developing Bayesian model-assisted inference under complex sampling is a challenging  problem in statistics.
%\cite{dong2014} developed a nonparametric method for Bayesian inference with survey data under the assumption that the sampling design is non-informative in the sense of \cite{pfeffermann1999}.
%This work extends the finite population
% Ghosh and Meeden (1983), and Cohen (1997) are based on nonparametric models.
%The idea of approximate Bayesian inference for developing a design-based posterior inference under %
\cite{wang2018}  recently propose the so-called approximate Bayesian method for design-based inference using asymptotic normality of a design-consistent  estimator. Specifically, for a given parameter $\theta$ with a prior distribution $\pi( \theta)$, if one can find a design-consistent estimator $\hat{\theta}$ of $\theta$, then the  approximate posterior distribution of $\theta$ is given by
\begin{equation}
 p( \theta \mid \hat{\theta} ) = \frac{ f ( \hat{\theta} \mid \theta ) \pi (\theta) }{ \int f ( \hat{\theta} \mid \theta ) \pi (\theta) {\rm d} \theta } ,
 \label{post}
\end{equation}
where $f ( \hat{\theta} \mid \theta )$ is the sampling distribution of $\hat{\theta}$, which is often approximated by a normal distribution.

Drawing on this idea, one can develop an approximate Bayesian approach to capture  the full uncertainty in the regression estimator.
Let 
 $$\bbeh = \left( \sum_{i \in A} \pi_i^{-1} \bx_{i} \bx_{i}^t \right)^{-1}  \sum_{i \in A} \pi_i^{-1} \bx_{i} y_i$$  be the design-consistent estimator of $\bbeta$   and $\bVh_\beta$ be the corresponding asymptotic variance-covariance matrix of $\bbeh$, given by
\begin{equation}
\bVh_{\beta}  =\left( \sum_{i \in A} \pi_i^{-1} \bx_{i} \bx_{i}^t \right)^{-1}  \left( \sum_{i \in A} \sum_{j \in A} \frac{ \Delta_{ij} }{ \pi_{ij} } \frac{\hat{e}_i \bx_{i} }{ \pi_i} \frac{  \hat{e}_j \bx_j^t}{ \pi_j } \right)  \left( \sum_{i \in A} \pi_i^{-1} \bx_{i} \bx_{i}^t \right)^{-1},
\label{9}
\end{equation}
where $\hat{e}_i = y_i - \bx_{i}^t \hat{\bbeta}$, $\Delta_{ij} = \pi_{ij} - \pi_i \pi_j$ and $\pi_{ij}$ is the joint inclusion probability of unit $i$ and $j$. Under some regularity conditions, as discussed in Chapter 2 of \cite{fuller2009},  we can establish
\begin{equation}
  \bVh_{\beta 11}^{-1/2}  \left( \hat{\bbeta}_1  - \bbeta_1 \right) \mid \bbeta \stackrel{\mathcal{L}}{ \longrightarrow }  N(0, I )
  \label{clt}
  \end{equation}
as $n \rightarrow \infty$, where $\bVh_{ \beta 11} $ is the submatrix of $\bVh_{\beta}$ with
\begin{equation} 
 \bVh_{\beta} = \begin{pmatrix}
\Vh_{\beta 00}  & \Vh_{\beta 01}  \\
\Vh_{\beta 10}  & \Vh_{\beta 11}
\end{pmatrix}.
\label{cov}
\end{equation}

Thus, using (\ref{post}) and (\ref{clt}), we can obtain the approximate posterior distribution of $\bbeta$ as
\begin{equation}\label{pos-beta}
p(\bbeta_1|\bbeh_1)
=\frac{\phi_p (\bbeh_1; \bbeta_1, \bVh_{\beta 11}  )\pi(\bbeta_1 )}{\int\phi_p(\bbeh_1; \bbeta_1, \bVh_{\beta 11} )\pi(\bbeta_1){\rm d}\bbeta_1},
\end{equation}
where $\phi_p$ denotes a $p$-dimensional multivariate normal density and $\pi(\bbeta_1)$ is a  prior distribution for $\bbeta_1$.

Now, we consider the conditional posterior distribution of $\bar{Y}$ for a given $\bbeta_1$. First,  define
$$
\hbarY ( \bbeta_1) =  \bar{\bX}_1^t  \bbeta_1  + \frac{1}{\hat{N}} \sum_{i \in A} \frac{1}{\pi_i} \left( y_i - \bx_{1i}^t {\bbeta}_1 \right).
$$
Note that  $\hbarY(\bbeta_1)$ is an approximately design-unbiased estimator of $\bar{Y}$,  regardless of $\bbeta_1$. Under some regularity conditions, we can show that $\hbarY( \bbeta_1)$ follows a normal distribution asymptotically.
Thus, we obtain
\begin{equation}
\frac{ \hbarY(\bbeta_1)-\barY }{
\sqrt{\Vh_{e}(\bbeta_1)}}  \mid \bar{Y}, \bbeta_1 \stackrel{\mathcal{L}}{ \longrightarrow }  N(0,1),
\label{clt2}
\end{equation}
where
\begin{equation}\label{Ve}
\Vh_{e} (\bbeta_1)=\frac1{N^2}\sum_{i\in A}\sum_{j\in A}\frac{\Delta_{ij}}{\pi_{ij}} \frac{1}{ \pi_i} \frac{1}{\pi_j} (y_i-\bx_{1i}^t\bbeta_1)(y_j-\bx_{1j}^t\bbeta_1),
\end{equation}
is a design consistent variance estimator of $\hbarY(\bbeta_1)$ for given $\bbeta_1$.
We then use $\phi(\hbarY(\bbeta_1); \barY, \Vh_{e} (\bbeta_1))$ as the density for the approximate sampling distribution of $\hbarY (\bbeta_1)$  in (\ref{clt2}), where $\phi(\cdot; \mu,\sigma^2)$ is the normal density function with mean $\mu$ and variance $\sigma^2$.
Thus, the approximate conditional posterior distribution of $\barY$ given $\bbeta$ can be  defined as
\begin{equation}\label{cpos-Y}
p(\barY|\hbarY(\bbeta_1),\bbeta_1)\propto \phi(\hbarY(\bbeta_1); \barY, \Vh_{e} (\bbeta_1))\pi(\barY \mid \bbeta_1),
\end{equation}
where $\pi(\barY \mid \bbeta_1)$ is a conditional  prior distribution of $\barY$ given $\bbeta_1$. Without extra assumptions, we can use a flat prior distribution for $\pi ( \bar{Y} \mid \bbeta_1)$. 

Therefore, combining (\ref{pos-beta}) and (\ref{cpos-Y}), the approximate posterior distribution of $\barY$ can be  obtained as
\begin{equation}\label{pos-Y}
\begin{split}
&p(\barY|\hbarY(\bbeta_1),\bbeh_1)\\
& \ \ 
=\frac
{\int \phi(\hbarY(\bbeta_1); \barY, \Vh_{e}(\bbeta_1))\phi_p(\bbeh_1; \bbeta_1, \bVh_{\beta 11} )\pi(\bbeta_1)\pi(\barY \mid \bbeta_1){\rm d}\bbeta_1}
{\iint \phi(\hbarY(\bbeta_1); \barY, \Vh_{e} (\bbeta_1))\phi_p(\bbeh_1; \bbeta_1, \bVh_{\beta 11} )\pi(\bbeta_1)\pi(\barY \mid \bbeta_1 ){\rm d}\bbeta_1 {\rm d}\barY}.
\end{split}
\end{equation}
Generating posterior samples from (\ref{pos-Y}) can be easily carried out via the following two steps:

\begin{itemize}
\item[1.]
Generate posterior sample $\bbeta_1^{\ast}$ of $\bbeta_1$ from (\ref{pos-beta}).

\item[2.]
Generate posterior sample of $\barY$ from the conditional posterior (\ref{cpos-Y}) given $\bbeta_1^{\ast}$.
\end{itemize}

Based on the approximate posterior samples of $\barY$, we can compute the posterior mean as a point estimator as well as credible intervals for uncertainty quantification for $\barY$ including the variability in estimating  $\bbeta_1$.

The following theorem presents an asymptotic property of the proposed approximate Bayesian method.

\begin{theorem}\label{thm:BvM1}
Under the regularity conditions described in the Supplementary Material, conditional on the full sample data,
\begin{equation}\label{PN1}
\sup_{\barY\in \Theta_Y}\Big|p(\barY|\hbarY(\bbeta_1),\bbeh_1)-\phi(\barY;\hbarY,\Vh_{e} )\Big|\to 0,
\end{equation}
in probability as $n\to\infty$ and $n/N\to f\in [0,1)$, where $\Theta_Y$ is some Borel set for $\barY$ and $p(\barY|\hbarY(\bbeta_1),\bbeh_1)$ is given in (\ref{pos-Y}).
\end{theorem}

Theorem \ref{thm:BvM1} is a special case of the Bernstein-von Mises theorem \citep[][Section 10.2]{VdV2000} in survey regression estimation, and its sketched proof  is given in the Supplementary Material.  The proof is not necessarily rigorous but contains enough details to deliver the main ideas. 
According to Theorem \ref{thm:BvM1}, the credible  interval for $\barY$ constructed from the approximated posterior distribution (\ref{pos-Y}) is asymptotically equivalent to the frequentist confidence interval based on the asymptotic normality of the common survey regression estimator.
Therefore, the proposed Bayesian method implements 
 the frequentist inference of the  
 survey regression estimator at least asymptotically.
%The consistency of the Bayesian point estimator (e.g. posterior mean) follows directly from (\ref{PN1}) since $\Vh_{e}(\bbeh)\to 0$ in probability as $n\to\infty$.

%----------------------------------------------------------------%
%       ABR with shrinkage prior
%----------------------------------------------------------------%
\section{Approximate Bayesian method with shrinkage priors}

We now consider the  case  when there are many  auxiliary variables  in applying regression estimation.
%This problem is addressed in \cite{Mc2017} and \cite{RS1997}.
When $p$ is large, it is desirable  to select a suitable subset of auxiliary variables that are associated with the response variable to avoid inefficient regression estimation due to irrelevant covariates.  

To deal with the problem in a Bayesian way, we may define the approximate posterior distribution of $\barY$ given $\bbe_1$ as similar to (\ref{pos-Y}). That is, we use the asymptotic distribution of the estimators $\bbeh_1$ of $\bbe_1$ and assign a shrinkage prior for $\bbe_1$. 
Let $\pi_{\la}(\bbe_1)$ be the shrinkage prior for $\bbe_1$ with a structural parameter $\lambda$ which might be multivariate.

Among the  several choices of shrinkage priors, we specifically consider two priors for $\bbe_1$: Laplace \citep{ PC2008} and horseshoe \citep{Cal2009, Cal2010}.
The Laplace prior is given by $\pi_\la(\bbe_1)\propto \exp(-\la\sum_{k=1}^p|\beta_k|)$, which is related to Lasso regression \citep{tib1996}, so that the proposed approximated Bayesian method  can be seen as the Bayesian version of a survey regression estimator with Lasso \citep{Mc2017}.
The horseshoe prior is a more advanced shrinkage prior of the form:
\begin{equation}\label{HS}
\pi_{\la}(\bbe_1)=\prod_{k=1}^p\int_0^{\infty}\phi(\beta_k; 0,\la^2u_k^2)\frac{2}{\pi(1+u_k^2)}{\rm d}u_k,
\end{equation}
where $\phi(\cdot; a,b)$ denotes the normal density function with mean $a$ and variance $b$.
It is known that the horseshoe prior enjoys more severe shrinkage for the zero elements of $\bbe_1$ than the Laplace prior, thus  allowing strong signals to remain large \citep{Cal2009}.

Similarly to (\ref{pos-beta}), we can develop a posterior distribution of $\bbeta_1$ using the shrinkage prior 
\begin{equation}\label{mar:beta}
p_{\la} (\bbeta_1| \bbeh_1)
=\frac{\phi(\bbeh_1;\bbe_1,\Vh_{\beta 11} )\pi_{\la}(\bbe_1)}{\int\phi(\bbeh_1;\bbe_1,\Vh_{\beta 11} )\pi_{\la}(\bbe_1){\rm d}\bbe_1},
\end{equation}
where $\Vh_{\beta 11} $ is the asymptotic variance-covariance matrix of $\bbeh_1$, defined in (\ref{cov}).  Once $\bbe_1$ are sampled from (\ref{mar:beta}), we can use the same posterior distribution of $\bar{Y}$ in 
  (\ref{cpos-Y})  for a given $\bbe$.

Therefore, the approximate posterior distribution of $\barY$ can be obtained as
\begin{equation}\label{pos-Y2}
\begin{split}
&p_{\lambda}(\barY|\hbarY(\bbeta_1),\bbeh_1)\\
&\ \ =\frac
{\int \phi(\hbarY(\bbeta_1); \barY, \Vh_{e}(\bbeta_1))\phi_p(\bbeh_1; \bbeta_1, \bVh_{\beta 11} )\pi_{\lambda}(\bbeta_1)\pi(\barY \mid \bbeta_1){\rm d}\bbeta_1}
{\iint \phi(\hbarY(\bbeta_1); \barY, \Vh_{e} (\bbeta_1))\phi_p(\bbeh_1; \bbeta_1, \bVh_{\beta 11} )\pi_{\lambda} (\bbeta_1)\pi(\barY \mid \bbeta_1 ){\rm d}\bbeta_1 {\rm d}\barY}.
\end{split}
\end{equation}
Generating posterior samples from (\ref{pos-Y2}) can be easily carried out via the following two steps:
\begin{itemize}
\item[1.] For a given $\lambda$, generate posterior sample $\bbeta_1^{\ast}$ of $\bbeta_1$ from $p_\lambda  (\barY|\hbarY(\bbeta_1),\bbeh_1)$ in  (\ref{mar:beta}).
\item[2.]
Generate posterior sample of $\barY$ from the conditional posterior (\ref{cpos-Y}) for given $\bbeta_1^{\ast}$.
\end{itemize}

\begin{remark}
Let $\hat{\beta}_0^{(R)} $ and $\bbeh_1^{(R)}$ be the estimator of $\beta_0$ and $\bbe_1$ defined as
\begin{equation}\label{r-est}
(\hat{\beta}_0^{(R)},\bbeh_1^{(R)})={\rm argmin}_{\beta_0,\beta_1} \left\{\sum_{i\in A}\frac1{\pi_i}(y_i-\beta_0-\bx_{1i}^t\bbe_1)^2+{\rm P}_\la(\bbe_1)\right\},
\end{equation}
where ${\rm P}(\bbe_1)=-2\log\pi_{\la}(\bbe_1)$ is the penalty (regularization) term for $\bbe_1$ induced from prior $\pi_{\la}(\bbe_1)$.
For example, the Laplace prior for $\pi_{\la}(\bbe_1)$ leads to the penalty term  ${\rm P}(\bbe_1)=2\la\sum_{k=1}^p|\beta_k|$, in which $\bbeh_1^{(R)}$ corresponds to the regularized estimator of $\bbe_1$ used in \cite{Mc2017}.
Since the exponential of $-\sum_{i\in A}\pi_i^{-1}(y_i-\beta_0-\bx_i^t\bbe_1)^2$ is close to the approximated likelihood $\phi_p((\hat{\beta}_0,\bbeh_1^t); (\beta_0,\bbeta_1^t), \bVh_{\beta} )$ used in the approximated Bayesian method when $n$ is large, the mode of the approximated posterior of $(\beta_0,\bbe_1^t)$ would be close to the frequentist estimator (\ref{r-est}) as well.
\end{remark}
%Therefore, the proposed Bayesian method would produce almost the same point estimates as (\ref{r-est}).

%
\begin{remark}
By the frequentist approach, $\la$ is often called the tuning parameter and can be selected via a data-dependent procedure such as cross validation as used in \cite{Mc2017}.
On the other hand, in the Bayesian approach, we assign a prior distribution on the hyperparameter $\la$ and consider integration with respect to the posterior distribution of $\la$, which means that uncertainty of the hyperparameter estimation can be taken into account.
Specifically, we assign a gamma prior for $\la^2$ as  the Laplace prior and a half-Cauchy prior for $\la$ as  the horseshoe prior (\ref{HS}). 
They  both  lead to familiar forms of full conditional posterior distributions of $\la$ or $\la^2$.
The details are given in the Supplementary Material.
\end{remark}

As in Section \ref{sec:ABR}, we obtain the following asymptotic properties of the proposed approximate Bayesian method.

\begin{theorem}\label{thm:BvM2}
Under the regularity conditions described in the  Supplementary Material, conditional on the full sample data,
\begin{equation}\label{PN2}
\sup_{\barY\in \Theta_Y}\Big|p_{\la} (\barY|\hbarY(\bbeta_1),\bbeh_1)-\phi(\barY;\hbarY(\bbeh_1^{(R)}),\Vh_{e}(\bbeh_1^{(R)} ))\Big|\to 0,
\end{equation}
in probability as $n\to\infty$ and $n/N\to f\in [0,1)$, where $\Theta_Y$ is some Borel set for $\barY$ and $p_{\la} (\barY|\hbarY(\bbeta_1),\bbeh_1)$ is given in (\ref{pos-Y2}).
\end{theorem}

The sketched proof is given in the Supplementary Material.
Theorem \ref{thm:BvM2} ensures that the proposed approximate Bayesian method is asymptotically equivalent to the frequentist version in which $\bbe_1$ is estimated by the regularized method with penalty corresponding to the shrinkage prior used in the Bayesian method.
Moreover, the proposed Bayesian method can be extended to  cases using general non-linear regression, as demonstrated in the next section.

%----------------------------------------------------------------%
%       Non-linear model
%----------------------------------------------------------------%
\section{An Extension to non-linear models}

The proposed Bayesian methods can be readily extended to work with non-linear regression. Some extensions of the regression estimator to nonlinear models are also considered  in \cite{wu2001}, \cite{breidt05}, and \cite{montanari2005}.

We consider a general working model for $y_i$ as $\E(y_i\mid \bx_i)=m(\bx_i;\bbeta)=m_i$ and $\Var(y_i\mid \bx_i)=\sigma^2a(m_i)$ for some known functions $m(\cdot;\cdot)$ and $a(\cdot)$.
The model-assisted regression estimator for $\barY$  with  $\bbeta$ known is then
$$
\hat{\bar{Y}}_{\rm reg, m}({\bbe})=\frac1N\left\{\sum_{i=1}^Nm(\bx_i;{\bbe} )+\sum_{i\in A}\frac1{\pi_i}\Big(y_i-m(\bx_i;{\bbe} )\Big)\right\},
$$
and its design-consistent variance estimator is obtained by
\begin{equation*}
\Vh_{e,m} ({\bbeta} )=\frac1{N^2}\sum_{i\in A}\sum_{j\in A}\frac{\Delta_{ij}}{\pi_{ij}} \frac{1}{ \pi_i} \frac{1}{\pi_j} \{ y_i- m( \bx_i; {\bbe} )  \}  \{ y_j-  m( \bx_j ; {\bbe}) \} ,
\end{equation*}
 which gives the approximate conditional posterior distribution
 of $\barY$ given $\bbe$.  That is,  similarly to (\ref{cpos-Y}), we can obtain
  \begin{equation}\label{cpos-Y2}
p(\barY| \hat{\bar{Y}}_{\rm reg, m}({\bbe}) ,\bbeta)\propto \phi( \hat{\bar{Y}}_{\rm reg, m}({\bbe}); \barY, \Vh_{e,m } (\bbeta))\pi(\barY \mid \bbeta).
\end{equation}
%where $\pi(\barY)$ is a prior distribution of $\barY$.

  To generate the posterior values of $\bbeta$, we first find a design-consistent estimator $\hat{\bbeta}$ of $\bbeta$.  Note that a consistent estimator $\hat{\bbeta}$ can be obtained by solving
$$
\hat{U}(\bbe)\equiv \sum_{i\in A}\pi_i^{-1}\{y_i-m (\bx_i; \bbeta) \} h(\bx_i; \bbe)=0,
$$
where $h(\bx_i; \bbe)=(\partial m_i/\partial\bbe)/a(m_i)$. For example, for binary $y_i$, we may use a logistic regression model with $m(\bx_i;\bbe)=\exp(\bx_i^t\bbe)/\{1+\exp(\bx_i^t\bbe)\}$ and $\Var(y_i)=m_i(1-m_i)$, which leads to $h(\bx_i;\bbe)=\bx_i$.

Under some regularity conditions, we can establish the asymptotic normality of $\bbeh$. That is,
$$
\bVh_{\beta}^{-1/2} (\bbeh-\bbe) \mid \bbeta  \stackrel{\mathcal{L}}{ \longrightarrow } N(0, I) ,
$$
where
$$
\bVh_{\beta}  =\left\{ \sum_{i \in A} \frac{1}{\pi_i} \hat{\bh}_i \dot{m} (\bx_i; \hat{\bbeta})^{t} \right\}^{-1}  \left( \sum_{i \in A} \sum_{j \in A} \frac{ \Delta_{ij} }{ \pi_{ij} } \frac{\hat{e}_i \hat{\bh}_i }{ \pi_i} \frac{  \hat{e}_j \hat{\bh}_j^t}{ \pi_j } \right)  \left\{\sum_{i \in A} \frac{1}{\pi_i} \hat{\bh}_i\dot{m} (\bx_i; \hat{\bbeta})^t \right\}^{-1},
$$
%$$
%\hat{V}_\beta ( \bbeta)  = \sum_{i \in A} \sum_{j \in A} \frac{ \Delta_{ij} }{ \pi_{ij} } \frac{ 1}{ \pi_i \pi_j}  \{y_i-m (\bx_i; \bbeta) \} \{y_j-m (\bx_j; \bbeta) \} \bh(\bx_i; \bbe) \bh^t(\bx_i; \bbe).
%$$
with $\hat{e}_i=y_i-m(\bx_i;\bbeh)$, $\hat{\bh}_i = h( \bx_i; \hat{\bbeta})$, and $\dot{m} (\bx; \bbeta) = \partial m( \bx; \bbeta)/ \partial \bbeta$.
Note that $\dot{m} (\bx; \bbeta) =m_i(1-m_i)\bx_i$ under a logistic regression  model.

Thus, the posterior distribution of $\bbeta$ given $\hat{\bbeta}$ can be obtained by
\begin{equation}
p( \bbeta \mid \hat{\bbeta} ) \propto \phi ( \bbeh \mid \bbeta, \bVh_\beta ) \pi (\bbeta) .
\label{24}
\end{equation}
We can use a shrinkage prior $\pi ( \bbeta)$ for $\bbeta$ in (\ref{24}) if necessary.   Once $\bbeta^*$ is generated from (\ref{24}), the posterior values of $\bar{Y}$ are generated from (\ref{cpos-Y2}) for a given $\bbeta^*$.

This formula enables us to define the approximate posterior distribution of $\bbe$ of the form (\ref{pos-beta}), so that  the approximate Bayesian inference for $\barY$ can be carried out in the same way as in the linear regression case.
Note that Theorem \ref{thm:BvM1} still holds under the general setup  as long as the regularity conditions given in the Supplementary Material are satisfied.

%----------------------------------------------------------------%
%       Simulation studies
%----------------------------------------------------------------%
\section{Simulation}\label{sec:sim}

We investigate the performance of the proposed approximate Bayesian methods against standard frequentist methods using two limited simulation studies. In the first simulation, we consider a linear regression model for a continuous $y$ variable. In the second simulation, we consider a binary $y$ and apply the logistic regression model for the non-linear regression estimation.

%\subsection{Simulation study: linear regression}
In the first simulation, we generate $x_i=(x_{i1},\ldots,x_{ip^{\ast}})^t$, $i=1,\ldots,N$, from a multivariate normal distribution with mean vector $(1,\ldots,1)^t$ and variance-covariance matrix $2R(0.2)$, where $p^* = 50$ and the $(i,j)$-th element of $R(\rho)$ is $\rho^{|i-j|}$.
The response variables $Y_i$ are generated from the following linear regression model:
$$
Y_i=\beta_0+\beta_1x_{i1}+\cdots+\beta_{p^{\ast}}x_{ip^{\ast}}+\ep_i, \ \ \ \ i=1,\ldots,N,
$$
where $N=10,000$, $\ep_i\sim N(0,2)$, $\beta_1=1$, $\beta_4=-0.5$, $\beta_7=1$, $\beta_{10}=-0.5$ and the other $\beta_k$'s are set to zero.
For the dimension of the auxiliary information, we consider four scenarios for $p$ of $20, 30, 40$ and $50$.
For each $p$, we assume that we can access only  $(x_{i1},\ldots,x_{ip})^t$ a subset of  the full information $(x_{i1},\ldots,x_{ip^{\ast}})^t$.
Note that for  all scenarios the  auxiliary variables significantly related with $Y_i$ are included, and so  only the amount of irrelevant information gets larger as $p$ gets larger.
We selected a sample size of $n=300$ from the finite population, using two sampling mechanism: (A) simple random sampling (SRS)  and (B) probability-proportional-to-size sampling (PPS) with size measure $z_i=\max\{\log(1+|Y_i+e_i|) ,1\}$ with $e_i\sim {\rm Exp}(2)$. 
The parameter of  interest is  $\barY=N^{-1}\sum_{i=1}^NY_i$. We assume that $\bar{X}_k=N^{-1}\sum_{i=1}^Nx_{ik}$ is known for all $k=1,\ldots,p$.

For the simulated dataset, we apply the proposed approximate Bayesian methods with the uniform prior $\pi(\bbe_1)\propto 1$, Laplace prior and horseshoe prior (\ref{HS}) for $\bbe_1$, which are denoted by AB, ABL and ABH, respectively.
For  all the Bayesian methods, we use $\pi(\barY)\propto 1$.
We generate 5,000 posterior samples of $\barY$ after discarding the first 500 samples  and compute  the posterior mean of $\barY$ as the point estimate.
As for the frequentist methods, we apply the original generalized regression estimator without variable selection (GREG) as well as the GREG method with Lasso regularization \citep[GREG-L;][]{Mc2017}, ridge estimation of $\bbe_1$ \citep[GREG-R;][]{RS1997} and forward variable selection (GREG-V) using adjusted coefficient of determination.
We also adopted the mixed modeling approach to the GREG estimation \citep[GREG-M;][]{PF2009} which is similar to GREG-R.
Moreover, the HT estimator is employed as a benchmark for efficiency comparison.
In GREG-L, the tuning parameter is selected via 10-fold cross validation, and we use the gamma prior ${\rm Ga}(\la_{\ast}^2,1)$ for $\lambda^2$ in ABL, where $\la_{\ast}$ is the selected value for $\lambda$ in GREG-L.
In ABH, we assign a  prior for the tuning parameter and generate posterior samples.
Based on $1,000$ replications, we calculate the square root of mean squared errors (RMSE) and bias of point estimators which are reported in Table \ref{tab:sim-C1}.
We also evaluated the performance of $95\%$ confidence (credible) intervals using coverage probabilities (CP) and the average length (AL), which are shown in Table \ref{tab:sim-C2}.

Table \ref{tab:sim-C1} shows that RMSE and bias of AB and GREG are almost identical, which is consistent with the fact that AB is a Bayesian version of GREG. 
Moreover, the results show that the existing shrinkage methods such as GREG-L and the proposed Bayesian methods ABL and ABH tend to produce smaller RMSEs and smaller absolute biases than GREG or AB as $p$ increases, which indicates the importance of suitable selection of auxiliary variables when $p$ is large. 
From Table \ref{tab:sim-C2}, it is observed that the CPs of GREG decreases as $p$ increases and are significantly smaller than the nominal level since GREG ignores the variability in estimating $\bbeta$ and the variability increases as $p$ increases.
On the other hand, the Bayesian version AB can take account of the variability estimating $\bbeta$ and the CPs are around the nominal level and ALs of AB are larger than those of GREG.
Although the performance of GREG-L is much better than GREG due to the shrinkage techniques, the CPs are not necessarily close to the nominal level. 
Note that GREG-M takes account of the variability estimating $\bbeta$, but not in other parameters, thereby the coverage performance is limited. 
It is also confirmed that the proposed ABH and ABL methods produce narrower intervals than AB.

%\subsection{Simulation study: logistic regression}
%
In the second simulation study, we consider the binary case for $y_i$ and apply the non-linear regression method discussed in Section 5.
The binary response variables $Y_i$ are generated from the following logistic regression model:
$$
Y_i\sim {\rm Ber}(\delta_i), \ \ \ \log\left(\frac{\delta_i}{1-\delta_i}\right)=\beta_0+\beta_1x_{i1}+\cdots+\beta_{p}x_{ip}, \ \ \ \ i=1,\ldots,N,
$$
where $\beta_0=-1$ and the other settings are the same as the linear regression case.
We selected a sample size of $n=300$ from the finite population, using two sampling mechanism: (A) simple random sampling and (B) probability-proportional-to-size sampling with size measure $z_i=\max\{\log(1+0.5Y_i+e_i) ,0.5\}$ with $e_i\sim {\rm Exp}(3)$. 
We again apply the three Bayesian methods and three frequents methods, GREG, GREG-L and GREG-R, based on a logistic regression model to obtain point estimates and confidence/credible intervals of the population mean $\barY=N^{-1}\sum_{i=1}^NY_i$.
The obtained RMSE and bias of point estimates and CP and AL of intervals based on 1,000 replications are reported in Tables \ref{tab:sim-B1} and \ref{tab:sim-B2}, respectively, which also shows again the superiority of the proposed Bayesian approach to the frequentist approach in terms of uncertainty quantification.

%   Table (linear regression: MSE)
\begin{table}[!htb]
\caption{Square root of Mean squared errors (RMSE) and bias of point estimators under $p\in \{20, 30, 40, 50\}$ in scenarios (A) and (B) with linear regression.
All values are multiplied by 100.
\label{tab:sim-C1}
}
\begin{center}
\begin{tabular}{cccccccccccccccc}
\hline
&&& \multicolumn{4}{c}{(A)} && \multicolumn{4}{c}{(B)}\\
& Method& & 20 & 30 & 40 & 50 &&  20 & 30 & 40 & 50\\
\hline
 & GREG &  & 11.7 & 11.8 & 12.0 & 12.3 &  & 11.4 & 11.8 & 12.1 & 12.3 \\
 & GREG-L &  & 11.7 & 11.7 & 11.7 & 11.8 &  & 11.1 & 11.1 & 11.1 & 11.1 \\
 & GREG-R &  & 11.8 & 11.9 & 12.1 & 12.4 &  & 11.4 & 11.6 & 11.8 & 12.0 \\
 & GREG-V &  & 11.6 & 11.7 & 11.8 & 12.0 &  & 11.3 & 11.5 & 11.8 & 12.0 \\
MSE & GREG-M &  & 11.7 & 11.8 & 12.0 & 12.3 &  & 11.4 & 11.8 & 12.1 & 12.3 \\
 & AB &  & 11.7 & 11.9 & 12.1 & 12.4 &  & 11.6 & 11.9 & 12.2 & 12.5 \\
 & ABL &  & 11.7 & 11.8 & 11.9 & 12.2 &  & 11.4 & 11.7 & 11.8 & 12.0 \\
 & ABH &  & 11.6 & 11.6 & 11.6 & 11.8 &  & 11.2 & 11.3 & 11.3 & 11.4 \\
 & HT &  & 17.5 & 17.5 & 17.5 & 17.5 &  & 14.8 & 14.8 & 14.8 & 14.8 \\
 \hline
 & GREG &  & 0.21 & 0.12 & 0.13 & 0.23 &  & 0.54 & 1.24 & 1.87 & 2.41 \\
 & GREG-L &  & 0.19 & 0.16 & 0.18 & 0.19 &  & 0.00 & 0.11 & 0.20 & 0.26 \\
 & GREG-R &  & 0.22 & 0.16 & 0.18 & 0.31 &  & 0.56 & 1.21 & 1.79 & 2.32 \\
 & GREG-V &  & 0.16 & 0.05 & 0.08 & 0.17 &  & 0.29 & 0.80 & 1.26 & 1.64 \\
Bias & GREG-M &  & 0.21 & 0.12 & 0.13 & 0.23 &  & 0.54 & 1.24 & 1.87 & 2.41 \\
 & AB &  & 0.19 & 0.10 & 0.11 & 0.22 &  & 0.60 & 1.28 & 1.92 & 2.44 \\
 & ABL &  & 0.19 & 0.11 & 0.11 & 0.21 &  & 0.49 & 1.06 & 1.55 & 1.95 \\
 & ABH &  & 0.16 & 0.12 & 0.11 & 0.17 &  & 0.06 & 0.29 & 0.51 & 0.71 \\
 & HT &  & 0.78 & 0.78 & 0.78 & 0.78 &  & -1.08 & -1.08 & -1.08 & -1.08 \\
\hline
\end{tabular}
\end{center}
\end{table}

%   Table (linear regression: interval)
\begin{table}[!htb]
\caption{Coverage probabilities (CP) and average lengths (AL) of $95\%$ confidence/credible intervals under $p\in \{20, 30, 40, 50\}$ in scenarios (A) and (B) with linear regression.
All values are multiplied by 100.
\label{tab:sim-C2}
}
\begin{center}
\begin{tabular}{cccccccccccccccc}
\hline
&&& \multicolumn{4}{c}{(A)} && \multicolumn{4}{c}{(B)}\\
&Method& & 20 & 30 & 40 & 50 &&  20 & 30 & 40 & 50\\
\hline
 & GREG &  & 92.8 & 92.8 & 92.7 & 89.9 &  & 94.2 & 92.1 & 92.1 & 90.1 \\
 & GREG-L &  & 93.5 & 93.4 & 93.2 & 93.3 &  & 94.5 & 94.8 & 94.4 & 94.8 \\
 & GREG-R &  & 93.0 & 92.4 & 91.8 & 90.0 &  & 93.3 & 92.4 & 91.9 & 90.4 \\
 & GREG-V &  & 93.6 & 93.7 & 93.3 & 91.4 &  & 94.1 & 93.8 & 92.5 & 91.2 \\
CP & GREG-M &  & 93.9 & 93.9 & 93.9 & 92.9 &  & 94.5 & 93.7 & 93.8 & 92.9 \\
 & AB &  & 95.3 & 94.8 & 94.9 & 94.2 &  & 95.1 & 94.8 & 94.9 & 95.2 \\
 & ABL &  & 95.2 & 94.6 & 94.8 & 94.5 &  & 95.3 & 95.3 & 95.1 & 94.9 \\
 & ABH &  & 94.8 & 95.0 & 95.0 & 94.7 &  & 95.4 & 95.9 & 95.1 & 95.5 \\
 & HT &  & 94.5 & 94.5 & 94.5 & 94.5 &  & 95.2 & 95.2 & 95.2 & 95.2 \\
 \hline
 & GREG &  & 43.1 & 42.3 & 41.5 & 40.7 &  & 43.1 & 42.3 & 41.5 & 40.7 \\
 & GREG-L &  & 43.8 & 43.7 & 43.6 & 43.5 &  & 43.3 & 43.1 & 42.9 & 42.8 \\
 & GREG-R &  & 43.2 & 42.5 & 41.9 & 41.4 &  & 42.8 & 42.0 & 41.3 & 40.7 \\
 & GREG-V &  & 43.4 & 42.8 & 42.2 & 41.6 &  & 43.4 & 42.9 & 42.3 & 41.8 \\
AL & GRREG-M &  & 44.2 & 44.2 & 44.3 & 44.4 &  & 44.3 & 44.4 & 44.6 & 44.8 \\
 & AB &  & 45.8 & 46.3 & 46.8 & 47.3 &  & 46.2 & 47.0 & 47.8 & 48.7 \\
 & ABL &  & 45.6 & 45.9 & 46.1 & 46.3 &  & 45.8 & 46.4 & 46.8 & 47.3 \\
 & ABH &  & 45.1 & 45.2 & 45.2 & 45.1 &  & 45.2 & 45.4 & 45.4 & 45.6 \\
 & HT &  & 66.4 & 66.4 & 66.4 & 66.4 &  & 59.1 & 59.1 & 59.1 & 59.1 \\
\hline
\end{tabular}
\end{center}
\end{table}

%   Table (logistic regression: MSE)
\begin{table}[!htb]
\caption{Square root of Mean squared errors (RMSE) and bias of point estimators under $p\in \{20, 30, 40, 50\}$ in scenarios (A) and (B) with logistic regression.
All values are multiplied by 100.
\label{tab:sim-B1}
} 
\begin{center}
\begin{tabular}{cccccccccccccccc}
\hline
&&& \multicolumn{4}{c}{(A)} && \multicolumn{4}{c}{(B)}\\
 & Method &  & 20 & 30 & 40 & 50 &  & 20 & 30 & 40 & 50 \\
 \hline
 & GR &  & 2.24 & 2.29 & 2.32 & 2.36 &  & 2.32 & 2.39 & 2.50 & 2.57 \\
 & GRL &  & 2.17 & 2.18 & 2.19 & 2.20 &  & 2.27 & 2.29 & 2.31 & 2.30 \\
 & GRR &  & 2.22 & 2.26 & 2.29 & 2.31 &  & 2.32 & 2.38 & 2.44 & 2.49 \\
RMSE & AB &  & 2.23 & 2.26 & 2.28 & 2.30 &  & 2.31 & 2.37 & 2.45 & 2.50 \\
 & ABL &  & 2.21 & 2.23 & 2.24 & 2.25 &  & 2.27 & 2.28 & 2.26 & 2.23 \\
 & ABH &  & 2.18 & 2.20 & 2.23 & 2.26 &  & 2.26 & 2.27 & 2.28 & 2.32 \\
 & HT &  & 2.80 & 2.80 & 2.80 & 2.80 &  & 2.83 & 2.83 & 2.83 & 2.83 \\
 \hline
 & GR &  & -0.10 & -0.12 & -0.12 & -0.11 &  & 0.10 & 0.18 & 0.31 & 0.43 \\
 & GRL &  & -0.11 & -0.11 & -0.10 & -0.11 &  & 0.03 & 0.05 & 0.07 & 0.08 \\
 & GRR &  & -0.11 & -0.12 & -0.12 & -0.12 &  & 0.07 & 0.13 & 0.20 & 0.27 \\
Bias & AB &  & -0.11 & -0.13 & -0.13 & -0.13 &  & 0.09 & 0.17 & 0.27 & 0.38 \\
 & ABL &  & -0.10 & -0.10 & -0.07 & -0.02 &  & 0.07 & 0.13 & 0.19 & 0.22 \\
 & ABH &  & -0.10 & -0.11 & -0.10 & -0.11 &  & 0.01 & 0.03 & 0.04 & 0.03 \\
 & HT &  & -0.15 & -0.15 & -0.15 & -0.15 &  & 0.07 & 0.07 & 0.07 & 0.07 \\
\hline
\end{tabular}
\end{center}
\end{table}

%   Table (logistic regression: interval)
\begin{table}[!htb]
\caption{Coverage probabilities (CP) and average lengths (AL) of $95\%$ credible/confidence intervals under $p\in \{20, 30, 40, 50\}$ in scenarios (A) and (B) with logistic regression.
All values are multiplied by 100.
\label{tab:sim-B2}
}
\begin{center}
\begin{tabular}{cccccccccccccccc}
\hline
&&& \multicolumn{4}{c}{(A)} && \multicolumn{4}{c}{(B)}\\
 & Method &  & 20 & 30 & 40 & 50 &  & 20 & 30 & 40 & 50 \\
 \hline
 & GR &  & 92.3 & 90.8 & 88.8 & 86.4 &  & 91.9 & 90.3 & 87.3 & 84.6 \\
 & GRL &  & 94.1 & 94.1 & 93.9 & 93.2 &  & 93.2 & 93.0 & 92.6 & 92.9 \\
 & GRR &  & 92.8 & 92.1 & 91.0 & 90.6 &  & 92.0 & 90.8 & 89.6 & 89.0 \\
CP & AB &  & 94.8 & 95.5 & 95.4 & 96.1 &  & 94.6 & 94.1 & 94.5 & 95.1 \\
 & ABL &  & 95.1 & 95.7 & 95.9 & 96.5 &  & 94.6 & 95.2 & 96.6 & 97.2 \\
 & ABH &  & 95.1 & 96.0 & 96.0 & 96.2 &  & 95.1 & 95.2 & 95.9 & 96.2 \\
 & HT &  & 95.3 & 95.3 & 95.3 & 95.3 &  & 94.5 & 94.5 & 94.5 & 94.5 \\
 \hline
 & GR &  & 8.02 & 7.80 & 7.56 & 7.30 &  & 8.20 & 7.95 & 7.69 & 7.39 \\
 & GRL &  & 8.21 & 8.17 & 8.14 & 8.11 &  & 8.42 & 8.37 & 8.33 & 8.30 \\
 & GRR &  & 8.15 & 7.99 & 7.88 & 7.79 &  & 8.34 & 8.17 & 8.04 & 7.94 \\
AL & AB &  & 8.74 & 8.90 & 9.10 & 9.42 &  & 9.05 & 9.27 & 9.59 & 10.10 \\
 & ABL &  & 8.79 & 8.99 & 9.24 & 9.55 &  & 9.07 & 9.31 & 9.61 & 9.99 \\
 & ABH &  & 8.76 & 8.96 & 9.18 & 9.45 &  & 9.02 & 9.22 & 9.46 & 9.75 \\
 & HT &  & 11.14 & 11.14 & 11.14 & 11.14 &  & 11.00 & 11.00 & 11.00 & 11.00 \\
\hline
\end{tabular}
\end{center}
\end{table}

%----------------------------------------------------------------%
%       Example
%----------------------------------------------------------------%
\section{Example}\label{sec:exm}
We applied the proposed methods to the synthetic income data  available from the \verb+sae+ package \citep{MM2015} in \verb+R+.
In the dataset, the normalized annual net income is observed for a certain number of individuals in each province of Spain.
The dataset contains 9 covariates; four indicators of the four groupings of ages ($16-24$, $25-49$, $50-64$ and $\geq 65$ denoted by \verb+ag1+$,\ldots,$\verb+ag4+, respectively), the indicator of having Spanish nationality \verb+na+, the indicators of education levels (primary education \verb+ed1+ and post-secondary education \verb+ed2+), and the indicators of two employment categories (employed \verb+em1+ and unemployed \verb+em2+).
We also adopted 13 interaction variables:  \verb+ag1*na+, \verb+ag2*na+, \verb+ag3*na+, \verb+ag4*na+, \verb+ag2*ed1+, \verb+ag3*ed1+, \verb+ag4*ed1+, \verb+ag1*em1+, \verb+ag2*em1+, \verb+ag3*em1+, \verb+ag4*em1+, \verb+ed1*em1+ and \verb+ed2*em1+, as auxiliary variables, thereby $p=22$ in this example. 
The dataset also contains information of survey weights, so that we used its inverse value as the sampling probability.
Since there is no information regarding the details of sampling mechanism, we approximate the joint inclusion probability as the product of two sampling probabilities. 
In this example, we focus on estimating average income in three provinces, Palencia, Segovia and Soria, where the number of sampled units are 72, 58 and 20, respectively.
The number of non-sampled units were around $10^6$.
It should be noted that the number of sample sizes are not so large compared with the number of auxiliary variables, especially in Soria.
Hence, the estimation error of regression coefficients would not be negligible and the proposed Bayesian methods would be appealing in this case.

In order to perform joint estimation and inference in the three provinces, we employed the following working model:
\begin{equation}\label{working}
y_i = \alpha + \sum_{h\in \{1, 2,3\}} x_{0i}^{(h)} \beta_{0}^{(h)}  + \bx_i^t \bbeta_1 + e_i,
\end{equation}
where $\alpha$ is an intercept term, $x_{0i}^{(h)}=1$ if $i$ belong to province $h$, where $h=1$ for Palencia, $h=2$ for Segovia, and $h=3$ for Soria, and $\bx_i$ is the vector of auxiliary variables with dimension $p=22$ (9 auxiliary variables and 13 interaction variables).
Here $y_i$ is the log-transformed net income and $e_i$ is the error term.

Under the working model (\ref{working}), the posterior distribution of $\bar{Y}_h$ is
$$
  p\{ \bar{Y}_h \mid \hat{\bar{Y}}_{h, \text{reg}} ( \beta_0^{(h)}, \bbeta_1),\beta_0^{(h)}, \bbeta_1 \} \propto \phi (  \hat{\bar{Y}}_{h, \text{reg}} ( \beta_0^{(h)}, \bbeta_1) \mid \bar{Y}_h, \hat{V}_{e,h} (\bbeta) ) \pi ( \bar{Y}_h ) ,
$$
where
$$
\hat{\bar{Y}}_{h, \text{reg}}=   \hat{\beta}_0^{(h)} + \bar{\bX}_h^t  \hat{\bbeta}_1  + \frac{1}{N_h} \sum_{i \in A_h} \frac{1}{\pi_i} \left(  y_i -  \hat{\beta}_0^{(h)} - \bx_i^t \hat{\bbeta}_1 \right) ,
$$
and
$$
\hat{V}_{e,h} ( \bbeta) = \frac{1}{N_h^2} \sum_{i \in A_h} \sum_{j \in A_h} \frac{ \Delta_{ij} }{ \pi_{ij} } \frac{1}{\pi_i } \frac{1}{ \pi_j} \left( y_i - \beta_0^{(h)} -  \bx_i^t \bbeta_1 \right) \left( y_j - \beta_0^{(h)} -  \bx_j^t \bbeta_1 \right).
$$
Based on the above formulas, we performed the proposed approximate Bayesian methods for $\bar{Y}_h$ for each $h$, and computed $95\%$ credible intervals for the log-transformed average income with 5000 posterior samples after discarding the first 500 samples as burn-in period.
We considered three types of priors for $\bbe_1$, flat, Laplace and horseshoe priors as considered in Section \ref{sec:sim}.
We also calculated $95\%$ confidence intervals of the log-transformed average income based on the two frequentist methods, GREG and GREG-L, using the working model (\ref{working}).
In applying GREG-L, the tuning parameter in the Lasso estimator was selected via 10 fold cross validation.

\begin{figure}[!htb]
\centering
\includegraphics[width=15cm,clip]{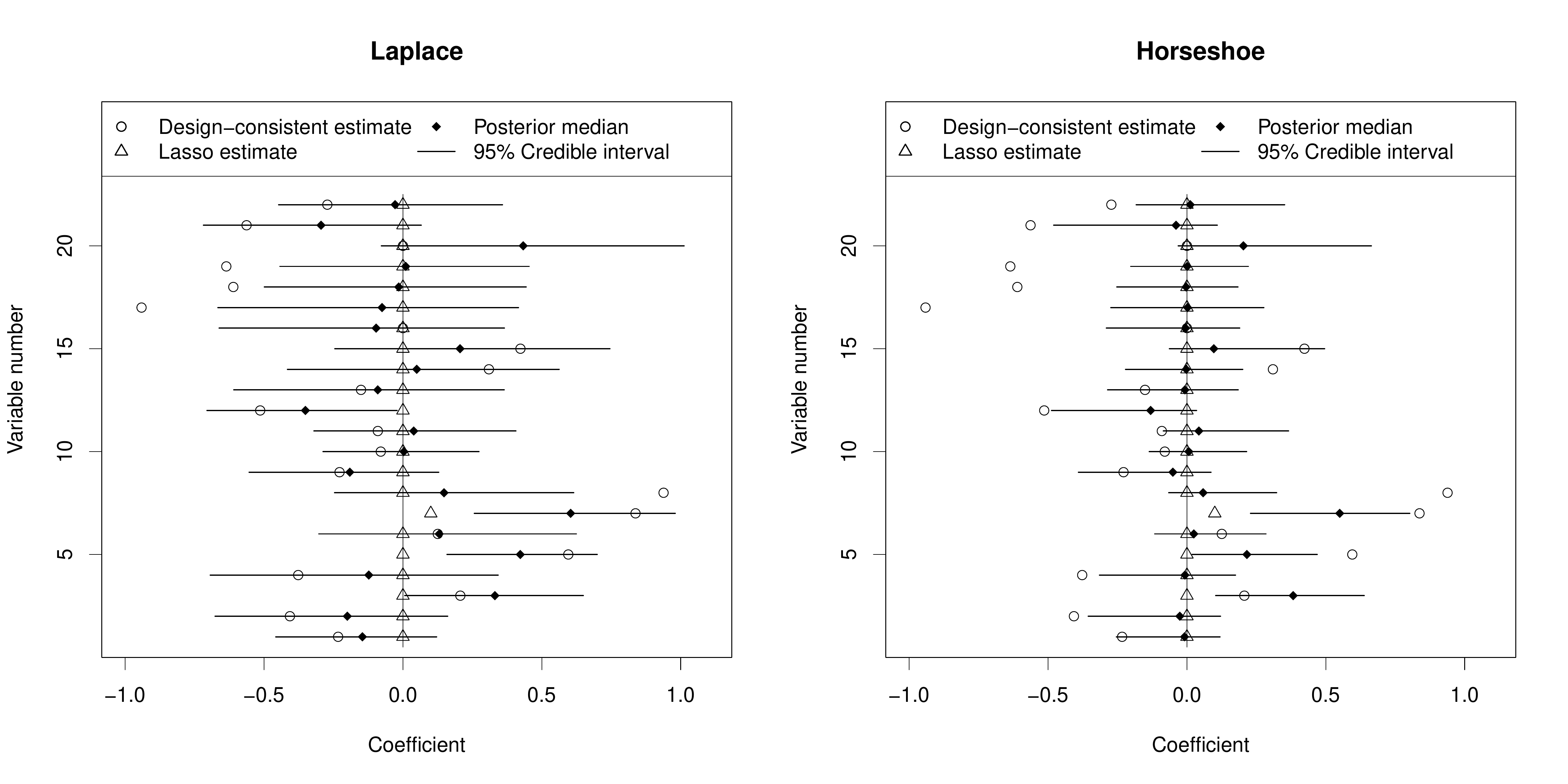}
\caption{
$95\%$ credible intervals of regression coefficients under Laplace (left) and horseshoe (right) priors.
}
\label{fig:beta}
\end{figure}

The $95\%$ credible intervals of $\bbe_1$ based on the approximate posterior distributions under Laplace and horseshoe priors are shown in Figure \ref{fig:beta}, in which the design-consistent and Lasso estimates of $\bbe_1$ are also given.
It is observed that the approximate posterior mean of $\bbe_1$ shrinks the design-consistent estimates of $\bbe_1$ toward $0$ although exactly zero estimates are not produced as  the frequentist Lasso estimator does.
The Lasso estimate selects only one variable among 22 candidates, and the variable is also significant in terms of the credible interval in both two priors.
Moreover, the two Bayesian methods detect one or two more variables to be significant judging from the credible intervals.
Comparing the results from two priors, the horseshoe prior provides narrower credible intervals than the Laplace prior.

\begin{figure}[!htb]
\centering
\includegraphics[width=12cm,clip]{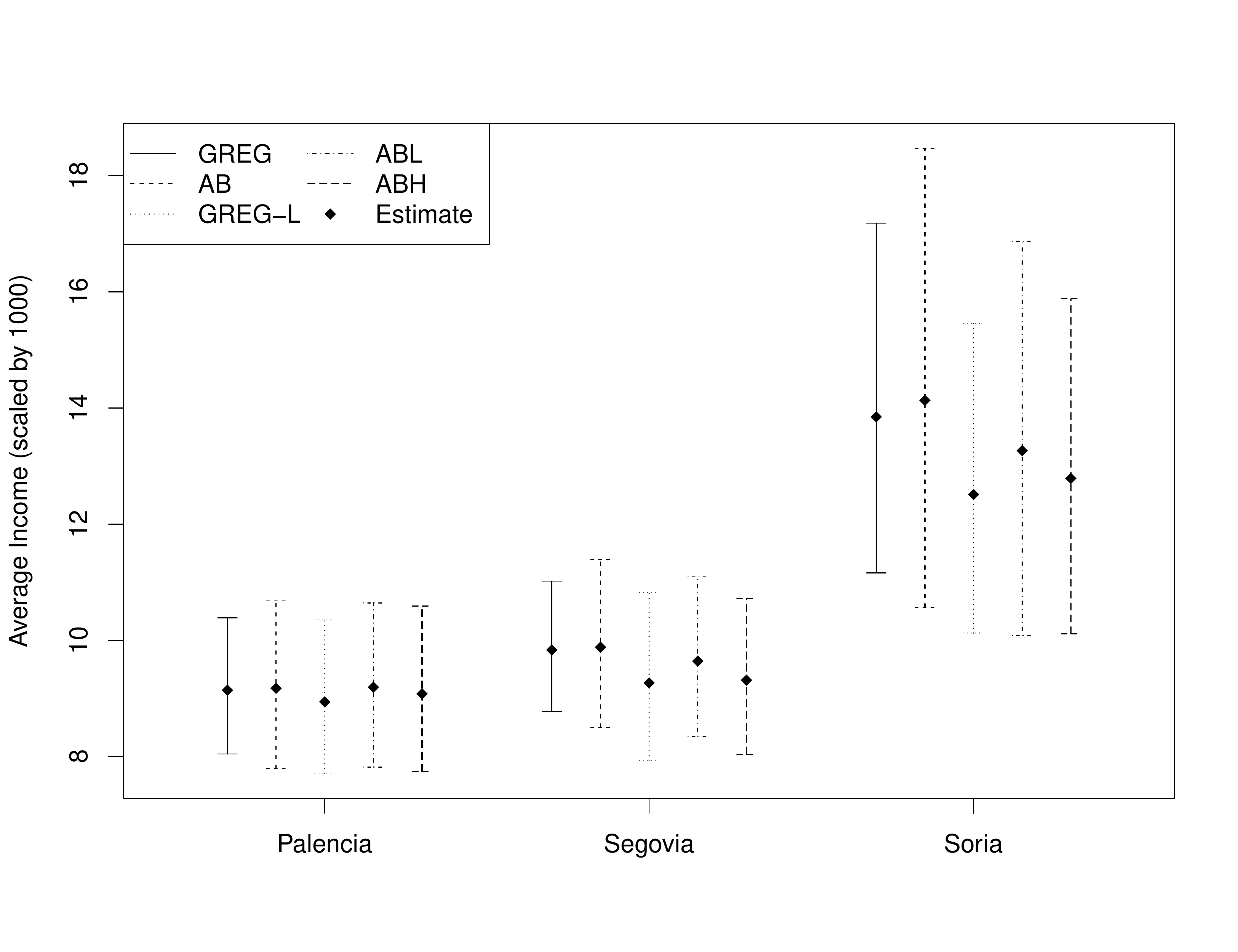}
\caption{$95\%$ confidence and credible intervals for average income based on five methods in three provinces in Spain.}
\label{fig:spain}
\end{figure}

In Figure \ref{fig:spain}, we show the resulting credible and confidence intervals of the average income in the three provinces.
It is observed that the proposed Bayesian methods, AB and ABL, tend to produce wider credible intervals than the confidence intervals of the corresponding frequencies methods, GREG and GREG-L, respectively, which is consistent with the simulation results in Section \ref{sec:sim}.
We can also confirm that the credible intervals of ABH are slightly narrower than those of ABL, which would reflect the differences of interval lengths of $\bbe_1$ as shown in Figure \ref{fig:beta}.

%----------------------------------------------------------------%
%       Discussion
%----------------------------------------------------------------%
\section{Concluding Remarks}
We have proposed an approximate Bayesian method for model-assisted survey estimation using parametric regression models as working models. The proposed  method is justified under the frequentist framework and 
captures the full uncertainty in  estimating regression parameters even when the number of the auxiliary variables is large. A main advantage of the proposed method is that it  uses a shrinkage prior for  regularized regression estimation, which not only provides an efficient point estimator, but also fully captures  the uncertainty associated with model selection and parameter estimation via  a Bayesian framework. Although we  only consider two popular prior distributions, the Laplace prior and the horseshoe prior,  other priors, such as the spike-and-slab prior \citep{ishwaran2005}, can be adopted in the same way.
Further investigation regarding the choice of the shrinkage prior distributions will be an important research topic in the future.
%The proposed method can be easily extended to non-linear models.
%Through simulation and empirical studies, the superiority of the proposed Bayesian methods to the classical inference methods has been confirmed.

%
Although our  working model is parametric, the proposed approximate Bayesian method  can be  applied to other semiparametric models such as local polynomial model \citep{breidt00}, P-spline regression model \citep{breidt05}, or a neural network model \citep{montanari2005}. By finding suitable prior distributions for the semiparametric models, the model complexity parameters will be determined automatically and the uncertainty will be captured in the approximate Bayesian framework.

%binary responses since the proposed method is based on the asymptotic normality of the design-consistent estimator of $\beta$ and the regression estimator $\hbarY$.
%However, the formula of asymptotic variance of $\hbarY$ might be different depending on the settings.
%The detailed investigation of the applications of the proposed methods to more complicated situations is left to a future study.

%
Finally, under more complicated sampling design such as multi-stage stratified cluster sampling, the main idea can be applied similarly since the proposed Bayesian method relies on the sampling distribution of the GREG estimator, which is asymptotically normal as shown by 
\cite{krewski1981}. If the asymptotic normality is questionable, one can use a weighted likelihood bootstrap to approximate Bayesian posterior, as in \cite{lyddon2019}. Such extensions are beyond the scope of this paper and will be considered in the future.

%----------------------------------------------------------------%
%       Supplement
%----------------------------------------------------------------%
\section*{Supplementary Materials}
Supplementary Material includes technical details for posterior computation, proofs of theorems and additional results of simulation studies.

%----------------------------------------------------------------%
%       Acknowledgement
%----------------------------------------------------------------%
\section*{Acknowledgement}
We thank the AE and three anonymous referees for very constructive comments. 
The first author was supported by Japan Society for the Promotion of Science KAKENHI grant number JP18K12757.
The second author was supported by US National Science Foundation (MMS-1733572).

\vspace{0.5cm}

\bibliographystyle{chicago}
\bibliography{ref}

%----------------------------------------------------------------%
%       Supplementary Material
%----------------------------------------------------------------%

%  new environment
\setcounter{equation}{0}
\renewcommand{\theequation}{S\arabic{equation}}
\setcounter{section}{0}
\renewcommand{\thesection}{S\arabic{section}}

\newpage
\begin{center}
{\Large
{\bf Supplementary Material for `An Approximate Bayesian Approach to Model-assisted Survey Estimation with Many Auxiliary Variables'}
}
\end{center}

\medskip
This Supplementary Material contains a proof of (5), details of posterior computation, proofs of theorems and results of additional simulation suites. 

\section{Proof of (2.5)}

We assume the same conditions in the proof of Theorem 1, given in Section \ref{sec:proof1}.
From (2.4), we have
\begin{eqnarray*}
E( R_n) & = & - E\left\{ ( \hat{\bar{\bX}}_{\rm HT} - \bar{\bX}_N )^t ( \bbeh_1 - \bbeta_{1\ast} ) \right\}
= - \mbox{tr} \left\{ {\rm Cov} \left( \hat{\bar{\bX}}_{\rm HT}, \bbeh_1 \right)  \right\} \\
&=& - \sum_{j=1}^p {\rm Cov} \left( \hat{\bar{x}}_{\rm HT, j}  , \hat{\beta}_j \right)= O(p/n),
\end{eqnarray*}
where the expectation is taken with respect to the sampling distribution. 
Also,
we can show that $ V( R_n ) =   O( p/ n^2 )$ .
Therefore, using Chebychev inequality, we have $R_n = O_p  ( p/n) $ and result (2.5) follows.

% Posterior computation
\section{Posterior computation}
We provide the algorithm for generating the approximate posterior distribution of $\bbe_1$ given in (4.20) with two shrinkage priors, Laplace and horseshoe (4.18) priors.
Using the mixture representation of both priors, we get the following Gibbs sampling algorithm.

\subsubsection*{Laplace prior}
We consider the mixture representation of Laplace distribution: $\beta_k|\tau_k\sim N(0,\tau_k^2)$ and $\tau_k^2\sim {\rm Exp}(\la^2/2)$, independently, for $k=1,\ldots,p$.
For $\lambda^2$, we consider the conjugate prior ${\rm Ga}(a,b)$, where ${\rm Ga}(a,b)$ is a gamma distribution with shape parameter $a$ and rate parameter $b$.
The full conditional distribution of $\bbeta_1$ is multivariate normal with mean $\bA^{-1}\bVh_{\beta11}^{-1}\bbeh_1$ and variance-covariance matrix $\bA^{-1}$ where $\bA=\bVh_{\beta11}^{-1}+\bD^{-1}$ with $\bD={\rm diag}(\tau_1^2,\ldots,\tau_p^2)$.
The full conditional distribution of $\lambda^2$ is ${\rm Ga}(a+p,b+\sum_{k=1}^p\tau_k^2/2)$, and $\tau_1^2,\ldots,\tau_p^2$ are conditionally independent, with $1/\tau_j^2$ conditionally inverse-Gaussian with parameters $\mu=\sqrt{\la/\beta_j^2}$ in the parametrization of the inverse-Gaussian density given by
$$
f(x)=\sqrt{\frac{\la}{2\pi}}x^{-3/2}\exp\left\{-\frac{\la(x-\mu)^2}{2\mu^2x}\right\}, \ \ x>0.
$$

\subsubsection*{Horseshoe prior}
The prior for $\bbe_1$ can be expressed as a hierarchy: $\beta_k|u_k\sim N(0,\la^2u_k^2)$ and $u_k\sim {\rm HC}(0,1)$ independently for $k=1,\ldots,p$, where ${\rm HC}(0,1)$ is the standard half-Cauchy distribution.
Using the hierarchical expression of the half-Cauchy distribution, we obtain the following Gibbs sampling steps.
Let $\bA=\bVh_{\beta11}^{-1}+\bB^{-1}$, where $\bB=\la^2{\rm diag}(u_1^2,\ldots,u_p^2)$.
The full conditional distribution of $\bbeta_1$ is multivariate normal with mean $\bA^{-1}\bVh_{\beta11}^{-1}\bbeh_1$ and variance-covariance matrix $\bA^{-1}$.
The full conditional distribution of $u_k^2$ and $\la^2$ are, respectively, give by
$$
{\rm IG}\left(1,\frac{1}{\xi_k}+\frac{\beta_k^2}{2\la^2}\right) \ \ \ \ \text{and} \ \ \ \
{\rm IG}\left(\frac{p+1}{2},\frac{1}{\gamma}+\frac12\sum_{k=1}^p\frac{\beta_k^2}{u_k^2}\right),
$$
where ${\rm IG}(a,b)$ denotes an inverse-Gamma distribution with shape parameter $a$ and rate parameter $b$.
Here $\xi_k$ and $\gamma$ are additional latent variables, and their full conditional distributions are given by ${\rm IG}(1,1+1/\delta_k^2)$ and ${\rm IG}(1,1+1/\la^2)$, respectively.

% Proof 1
\section{A sketched proof of Theorem 1}\label{sec:proof1}
To discuss the asymptotic properties of  the approximate Bayesian method, we first assume a sequence of finite populations and samples with finite fourth moments as in \cite{isaki1982}.
The finite population is a random sample from an unknown superpopulation model.
Let $\barY_{\ast}$ and $\bbe_{1\ast}$ be the true values of $\barY$ and $\bbe_1$.
Let $B_n=(\barY_{\ast}-r_n,\barY_{\ast}+r_n)$ and $C_n$ be a  ball with centre $\bbe_{1\ast}$ and radius $r_n\sim n^{\tau-1/2}$ for $0<\tau<1/2$.
We make the following regularity assumptions

\begin{itemize}
\item[(C1)]
Assume that the sufficient conditions for the asymptotic normality of $\hbarY$ for $\barY\in B_n$ hold for the sequence of finite populations and samples.

\item[(C2)]
Assume that the prior distribution $\pi(\barY)$ is positive and satisfies a Lipschitz condition over its support $\Theta_Y$; that is, there exists $C_1<\infty$ such that $|\pi(\theta_1)-\pi(\theta_2)|\leq C_1|\theta_1-\theta_2|$ for $\theta_1,\theta_2\in \Theta_Y$.

\item[(C3)]
Assume that $\bVh_{\beta11}=\bV_{\beta11}\{1+o_P(1)\}$ and $(\bbeh_1-\bbe_1)^t\bVh_{\beta11}^{-1}(\bbeh_1-\bbe_1)=(\bbeh_1-\bbe_1)^t\bV_{\beta11}^{-1}(\bbeh_1-\bbe_1)\{1+o_P(1)\}$ for any $\bbe\in C_n$ and $n\to\infty$.

\item[(C4)]
Assume that $\pi(\bbe)$ is positive and finite over its support $\Theta_\beta$.

\end{itemize}

Sufficient conditions for (C1) are discussed within various asymptotic structures \citep[e.g.][]{Binder1983, PS2009}.
Conditions (C2) and (C4) are satisfied for common priors such as (multivariate) normal distribution .
Condition (C3) essentially requires that the design variance estimators be consistent and meet a certain continuity condition.

\bigskip
\begin{proof}
Let $g(\barY,\bbe)=\phi(\hbarY(\bbeta_1); \barY, \Vh_{e}(\bbe))\phi_p(\bbeh_1; \bbeta_1, \bVh_{\beta11})\pi(\bbeta_1)$.
Then, the approximated posterior distribution is given by
\begin{align*}
p(\barY|\hbarY(\bbeh_1),\bbeh_1)
&=\frac{\int g(\barY,\bbe_1){\rm d}\bbe_1}{\iint g(\barY,\bbe_1){\rm d}\bbe_1{\rm d}\barY}.
\end{align*}
Note that
\begin{align} \label{eval.g}
\int g(\barY,\bbe_1){\rm d}\bbe_1
=\int_{\beta\in C_n}g(\barY,\bbe_1){\rm d}\bbe_1+\int_{\beta\in \mathbb{R}^p\setminus C_n}g(\barY,\bbe_1){\rm d}\bbe_1
\end{align}
By  the same argument in the proof of Theorem 1 in \cite{wang2018}, we have
$$
\plim_{n\to\infty}\int_{\beta\in C_n}\phi_p(\bbeh_1; \bbeta_1, \bVh_{\beta11}){\rm d}\bbe_1=1,
$$
so the second term in (\ref{eval.g}) is $o_P(1)$.
On the other hand, under condition (C3), $\phi_p(\bbeh_1;\bbe_1,\bVh_{\beta11})=\phi_p(\bbeh_1; \bbe_1,\bV_{\beta11})\{1+o_P(1)\}$ as $n\to\infty$, for any $\bbe_1\in C_n$, thereby under condition (C4),
\begin{align*}
\int_{\beta\in C_n}g(\barY,\bbe_1){\rm d}\bbe_1
&=\int_{\beta\in C_n}\phi(\hbarY(\bbeta_1); \barY, \Vh_{e}(\bbe_1))\phi_p(\bbeh_1; \bbeta_1, \bV_{\beta11})\pi(\bbeta_1){\rm d}\bbe_1\\
&=\phi(\hbarY(\bbe_{1\ast}); \barY, \Vh_{e}(\bbe_{1\ast}))\pi(\bbe_{1\ast})\{1+o_P(1)\}
\end{align*}
as $n\to\infty$ since $V\to 0$ and $\bbeh_1\to\bbe_{1\ast}$ as $n\to\infty$.
Hence, we have
\begin{align}
p(\barY|\hbarY(\bbeh_1),\bbeh_1)
&=\frac
{\pi(\bbe_{1\ast})\phi(\hbarY(\bbe_{1\ast}); \barY, \Vh_{e}(\bbe_{1\ast}))\pi(\barY)\{1+o_P(1)\}}
{\pi(\bbe_{1\ast})\int\phi(\hbarY(\bbe_{1\ast}); \barY, \Vh_{e}(\bbe_{1\ast}))\pi(\barY)d\barY\{1+o_P(1)\}} \notag\\
&=\frac{\pi(\barY)}{\pi(\barY_{\ast})}
\phi(\hbarY(\bbe_{1\ast}); \barY, \Vh_{e}(\bbe_{1\ast}))\{1+o_P(1)\}\notag\\
&=\phi(\hbarY(\bbe_{1\ast}); \barY, \Vh_{e}(\bbe_{1\ast}))\{1+o_P(1)\} \label{approx1}\\
&=\phi(\hbarY(\bbeh_1); \barY, \Vh_{e}(\bbeh_1))\{1+o_P(1)\},  \label{approx2}
\end{align}
for any $\barY\in B_n$ as $n\to\infty$, where (\ref{approx1}) follows from (C2), and (\ref{approx2}) follows from the properties $\Vh_{e}(\bbeh_1)=\Vh_{e}(\bbe_{1\ast})\{1+o_P(1)\}$ and $\hbarY(\bbeh_1)=\hbarY(\bbe_{1\ast})\{1+o_P(1)\}$ under (C1).
Let $R_n=\{\barY\in\Theta_Y : \Vh_{e}(\bbeh_1)^{-1}(\hbarY(\bbeh_1)-\barY)^2\leq \chi^2_1(q)\}$, where $\chi^2_k(q)$ is the upper $100q\%$-quantile of the chi-squared distribution with $k$ degree of freedom.
Then, $\plim_{n\to\infty}P(R_n)=q$.
Since $\hbarY(\bbeh_1)-\barY_{\ast}=O_p(n^{-1/2})$ and $r_n=n^{\tau-1/2}$, which is slower than $n^{-1/2}$, it holds that $\lim_{n\to\infty} P(R_n\subset B_n)=1$.
Then,
\begin{align*}
&\lim_{n\to\infty}P\left(\int_{B_n}\phi(\hbarY(\bbeh_1); \barY, \Vh_{e}(\bbeh_1)){\rm d}\barY \geq \int_{R_n}\phi(\hbarY(\bbeh_1); \barY, \Vh_{e}(\bbeh_1)){\rm d}\barY\right)=1,
\end{align*}
which means that
\begin{align*}
\lim_{n\to\infty}P\left(\int_{B_n}\phi(\hbarY(\bbeh_1); \barY, \Vh_{e}(\bbeh_1)){\rm d}\barY \geq q\right)=1
\end{align*}
for any $q\in (0,1)$, implying 
\begin{equation}\label{post.int}
\plim_{n\to\infty}\int_{B_n}\phi(\hbarY(\bbeh_1); \barY, \Vh_{e}(\bbeh_1)){\rm d}\barY=1.
\end{equation}
Then,
\begin{align*}
&\sup_{\barY\in \Theta_Y}\Big|p(\barY|\hbarY(\bbeh_1),\bbeh_1)-\phi(\barY;\hbarY(\bbeh_1),\Vh_{e}(\bbeh_1))\Big|\\
\leq &
\sup_{\barY\in B_n}\Big|p(\barY|\hbarY(\bbeh_1),\bbeh_1)-\phi(\barY;\hbarY(\bbeh_1),\Vh_{e}(\bbeh_1))\Big|\\
& +
\sup_{\barY\in \Theta_Y\setminus B_n}\Big|p(\barY|\hbarY(\bbeh_1),\bbeh_1)-\phi(\barY;\hbarY(\bbeh_1),\Vh_{e}(\bbeh_1))\Big|,
\end{align*}
which are both $o_P(1)$ from (\ref{approx2}) and (\ref{post.int}).  This  completes the proof.
\end{proof}

% Proof 2
\section{A sketched proof of Theorem 2}
The condition (C4) given in the proof of Theorem 1 may not be satisfied for shrinkage priors.
For example, the horseshoe prior diverge at the origin $\beta_k=0$.
In what follows, let $\bbe=(\beta_0,\bbe_1^t)$ and define $\bbeh_1$ and $\bbeh_1^{(R)}$ in the same way.
We use the following alternative condition for the shrinkage prior $\pi_{\la}(\bbe_1)$:
\begin{itemize}
\item[(C5)]
The regularized estimator $\bbeh_1^{(R)}$ under penalty $-\log\pi_{\la}(\bbe_1)$ is asymptotically normal, that is, $\sqrt{n}(\bbeh_1^{(R)}-\bbe_{1\ast})\to N(0,\bC)$, where $\bC$ is a positive definite matrix and $\la$ is appropriately chosen.
\end{itemize}

Under the Laplace prior, $\bbeh_1^{(R)}$ is equivalent to the Lasso estimator, and the above property holds if $\lambda=o(\sqrt{n})$ \citep{KF2000, Mc2017}.
For general prior $\pi_\la(\bbe_1)$, this condition holds if the assumption regarding the penalty term $P_\la(\bbe_1)$ given in \cite{FL2001} is satisfied.

\bigskip
\begin{proof}
It is noted that
\begin{align*}
\phi_p&((\hat{\beta}_0,\bbeh_1^t); (\beta_0,\bbeta_1^t), \bVh_{\beta11})\pi_\la(\bbe_1)\\
&\propto \exp\left\{-\frac12(\bbeh_1-\bbe)^t\bVh_{\beta11}^{-1}(\bbeh_1-\bbe)+\log\pi_\la(\bbe_1)\right\}\\
&=\exp\left\{-\frac12\sum_{i\in A}\frac1{\pi_i}(y_i-\beta_0-x_i^t\bbe_1)^2+\log\pi_\la(\bbe_1)\right\}\{1+o_P(1)\}\\
&=\exp\left\{-\frac{n}2(\bbeh_1^{(R)}-\bbe_1)^t\bC^{-1}(\bbeh_1^{(R)}-\bbe_1)\right\}\{1+o_P(1)\}.
\end{align*}
Define
$$
g(\barY,\bbe_1)=\phi(\hbarY(\bbeta_1); \barY, \Vh_{e}(\bbe_1))\phi(\bbeh_1;\bbe_1,\bVh_{\beta11})\pi_{\la}(\bbe_1).
$$
Then, it holds that
\begin{align*}
\int_{\bbe_1\in R_n}g(\barY,\bbe_1){\rm d}\bbe_1
=\phi(\hbarY(\bbe_{1\ast}); \barY, \Vh_{e}(\bbe_{1\ast})) \{1+o_P(1)\}
\end{align*}
as $n\to\infty$, where $R_n$ is a  ball with center $\bbe_{1\ast}$ and radius $O(n^{\tau-1/2})$ for $0<\tau<1/2$.
Hence, the statement can be proved in the same way as the proof of Theorem 1 since $\phi(\hbarY(\bbe_{1\ast}); \barY, \Vh_{e}(\bbe_{1\ast}))=\phi(\hbarY(\bbeh_1^{(R)}); \barY, \Vh_{e}(\bbeh_1^{(R)}))\{1+o_P(1)\}$.
\end{proof}

\section{Additional simulation results}
We here provide additional simulation results.
We considered the same scenarios in the main document with $n=400$.
The results are reported in Table S1$\sim$4.

%   Table (linear regression: MSE)
\begin{table}[!htb]
\caption{Square root of Mean squared errors (RMSE) and bias of point estimators under $p\in \{20, 30, 40, 50\}$ in scenarios (A) and (B) with linear regression.
All values are multiplied by 100.
\label{tab:sim-C1}
}
\begin{center}
\begin{tabular}{cccccccccccccccc}
\hline
&&& \multicolumn{4}{c}{(A)} && \multicolumn{4}{c}{(B)}\\
& Method& & 20 & 30 & 40 & 50 &&  20 & 30 & 40 & 50\\
\hline
 & GREG &  & 10.3 & 10.4 & 10.6 & 10.9 &  & 9.9 & 10.1 & 10.3 & 10.5 \\
 & GREG-L &  & 10.2 & 10.2 & 10.2 & 10.3 &  & 9.6 & 9.6 & 9.6 & 9.6 \\
 & GREG-R &  & 10.3 & 10.5 & 10.7 & 10.9 &  & 9.8 & 10.0 & 10.2 & 10.3 \\
 & GREG-V &  & 10.2 & 10.4 & 10.5 & 10.7 &  & 9.8 & 9.9 & 10.1 & 10.2 \\
RMSE & GREG-M &  & 10.3 & 10.4 & 10.6 & 10.9 &  & 9.9 & 10.1 & 10.3 & 10.5 \\
 & AB &  & 10.3 & 10.5 & 10.7 & 11.0 &  & 10.0 & 10.2 & 10.5 & 10.6 \\
 & ABL &  & 10.3 & 10.4 & 10.6 & 10.8 &  & 9.9 & 10.0 & 10.2 & 10.3 \\
 & ABH &  & 10.2 & 10.2 & 10.3 & 10.3 &  & 9.7 & 9.7 & 9.8 & 9.8 \\
 & HT &  & 14.8 & 14.8 & 14.8 & 14.8 &  & 12.6 & 12.6 & 12.6 & 12.6 \\
 \hline
 & GREG &  & -0.15 & -0.13 & -0.18 & -0.22 &  & 0.43 & 0.93 & 1.39 & 1.86 \\
 & GREG-L &  & -0.22 & -0.22 & -0.25 & -0.24 &  & 0.12 & 0.18 & 0.25 & 0.34 \\
 & GREG-R &  & -0.18 & -0.18 & -0.22 & -0.26 &  & 0.44 & 0.94 & 1.39 & 1.87 \\
 & GREG-V &  & -0.21 & -0.22 & -0.24 & -0.23 &  & 0.27 & 0.62 & 0.95 & 1.30 \\
Bias & GREG-M &  & -0.15 & -0.13 & -0.18 & -0.22 &  & 0.43 & 0.93 & 1.39 & 1.86 \\
 & AB &  & -0.17 & -0.16 & -0.20 & -0.24 &  & 0.43 & 0.93 & 1.37 & 1.85 \\
 & ABL &  & -0.19 & -0.18 & -0.22 & -0.25 &  & 0.38 & 0.80 & 1.17 & 1.56 \\
 & ABH &  & -0.20 & -0.20 & -0.21 & -0.25 &  & 0.14 & 0.30 & 0.46 & 0.63 \\
 & HT &  & -0.29 & -0.29 & -0.29 & -0.29 &  & -0.39 & -0.39 & -0.39 & -0.39 \\
\hline
\end{tabular}
\end{center}

\end{table}

%   Table (linear regression: interval)
\begin{table}[!htb]
\caption{Coverage probabilities (CP) and average lengths (AL) of $95\%$ confidence/credible intervals under $p\in \{20, 30, 40, 50\}$ in scenarios (A) and (B) with linear regression.
All values are multiplied by 100.
\label{tab:sim-C2}
}
\begin{center}
\begin{tabular}{cccccccccccccccc}
\hline
&&& \multicolumn{4}{c}{(A)} && \multicolumn{4}{c}{(B)}\\
&Method& & 20 & 30 & 40 & 50 &&  20 & 30 & 40 & 50\\
\hline
 & GREG &  & 93.8 & 92.4 & 91.1 & 89.4 &  & 94.0 & 92.6 & 91.5 & 91.4 \\
 & GREG-L &  & 94.2 & 94.0 & 94.1 & 93.9 &  & 94.9 & 94.4 & 94.8 & 94.5 \\
 & GREG-R &  & 93.3 & 92.2 & 91.8 & 90.1 &  & 94.4 & 93.0 & 91.7 & 91.7 \\
 & GREG-V &  & 94.1 & 93.3 & 92.8 & 90.9 &  & 94.5 & 94.0 & 92.8 & 92.4 \\
CP & GREG-M &  & 94.0 & 93.0 & 92.8 & 91.6 &  & 94.9 & 93.7 & 93.5 & 93.5 \\
 & AB &  & 94.2 & 94.2 & 94.5 & 94.4 &  & 95.6 & 95.5 & 94.3 & 94.7 \\
 & ABL &  & 94.2 & 94.3 & 94.9 & 94.9 &  & 95.6 & 95.0 & 94.5 & 94.6 \\
 & ABH &  & 94.6 & 94.8 & 94.6 & 94.4 &  & 95.2 & 95.3 & 95.5 & 95.5 \\
 & HT &  & 94.2 & 94.2 & 94.2 & 94.2 &  & 94.8 & 94.8 & 94.8 & 94.8 \\
 \hline
 & GREG &  & 37.4 & 36.9 & 36.4 & 35.9 &  & 37.6 & 37.1 & 36.6 & 36.1 \\
 & GREG-L &  & 37.9 & 37.8 & 37.7 & 37.7 &  & 37.7 & 37.6 & 37.5 & 37.4 \\
 & GREG-R &  & 37.5 & 37.0 & 36.6 & 36.2 &  & 37.4 & 36.9 & 36.4 & 35.9 \\
 & GREG-V &  & 37.6 & 37.2 & 36.8 & 36.4 &  & 37.8 & 37.5 & 37.2 & 36.8 \\
AL & GREG-M &  & 38.1 & 38.1 & 38.2 & 38.2 &  & 38.4 & 38.5 & 38.6 & 38.7 \\
 & AB &  & 39.2 & 39.5 & 39.9 & 40.2 &  & 39.6 & 40.2 & 40.8 & 41.3 \\
 & ABL &  & 39.0 & 39.2 & 39.5 & 39.6 &  & 39.4 & 39.8 & 40.2 & 40.5 \\
 & ABH &  & 38.8 & 38.8 & 38.8 & 38.9 &  & 39.0 & 39.2 & 39.3 & 39.3 \\
 & HT &  & 57.5 & 57.5 & 57.5 & 57.5 &  & 51.1 & 51.1 & 51.1 & 51.1 \\
\hline
\end{tabular}
\end{center}
\end{table}

%   Table (logistic regression: MSE)
\begin{table}[!htb]
\caption{Square root of Mean squared errors (RMSE) and bias of point estimators under $p\in \{20, 30, 40, 50\}$ in scenarios (A) and (B) with logistic regression.
All values are multiplied by 100.
\label{tab:sim-B1}
}
\begin{center}
\begin{tabular}{cccccccccccccccc}
\hline
&&& \multicolumn{4}{c}{(A)} && \multicolumn{4}{c}{(B)}\\
 & Method &  & 20 & 30 & 40 & 50 &  & 20 & 30 & 40 & 50 \\
 \hline
 & GREG &  & 1.90 & 1.91 & 1.93 & 1.97 &  & 1.94 & 1.98 & 2.00 & 2.06 \\
 & GREG-L &  & 1.86 & 1.87 & 1.87 & 1.87 &  & 1.90 & 1.91 & 1.91 & 1.92 \\
 & GREG-R &  & 1.88 & 1.89 & 1.91 & 1.93 &  & 1.93 & 1.97 & 1.98 & 2.02 \\
RMSE & AB &  & 1.89 & 1.89 & 1.91 & 1.93 &  & 1.93 & 1.96 & 1.98 & 2.01 \\
 & ABL &  & 1.88 & 1.88 & 1.89 & 1.90 &  & 1.91 & 1.91 & 1.89 & 1.88 \\
 & ABH &  & 1.87 & 1.87 & 1.88 & 1.89 &  & 1.88 & 1.87 & 1.86 & 1.87 \\
 & HT &  & 2.36 & 2.36 & 2.36 & 2.36 &  & 2.39 & 2.39 & 2.39 & 2.39 \\
 \hline
 & GREG &  & -0.03 & -0.03 & -0.03 & -0.05 &  & -0.05 & 0.01 & 0.08 & 0.19 \\
 & GREG-L &  & -0.02 & -0.01 & -0.01 & -0.02 &  & -0.12 & -0.11 & -0.10 & -0.09 \\
 & GREG-R &  & -0.03 & -0.02 & -0.02 & -0.04 &  & -0.07 & -0.02 & 0.02 & 0.09 \\
Bias & AB &  & -0.04 & -0.03 & -0.03 & -0.06 &  & -0.06 & 0.00 & 0.06 & 0.15 \\
 & ABL &  & -0.03 & -0.02 & -0.01 & 0.00 &  & -0.07 & -0.02 & 0.02 & 0.08 \\
 & ABH &  & -0.02 & -0.02 & -0.02 & -0.02 &  & -0.11 & -0.11 & -0.11 & -0.11 \\
 & HT &  & 0.01 & 0.01 & 0.01 & 0.01 &  & -0.12 & -0.12 & -0.12 & -0.12 \\
\hline
\end{tabular}
\end{center}
\end{table}

%   Table (logistic regression: interval)
\begin{table}[!htb]
\caption{Coverage probabilities (CP) and average lengths (AL) of $95\%$ credible/confidence intervals under $p\in \{20, 30, 40, 50\}$ in scenarios (A) and (B) with logistic regression.
All values are multiplied by 100.
\label{tab:sim-B2}
} 
\begin{center}
\begin{tabular}{cccccccccccccccc}
\hline
&&& \multicolumn{4}{c}{(A)} && \multicolumn{4}{c}{(B)}\\
 & Method &  & 20 & 30 & 40 & 50 &  & 20 & 30 & 40 & 50 \\
 \hline
  & GREG &  & 92.8 & 91.2 & 90.6 & 89.5 &  & 92.9 & 91.8 & 91.5 & 89.6 \\
 & GREG-L &  & 93.0 & 92.9 & 92.9 & 93.0 &  & 94.3 & 94.3 & 94.5 & 94.0 \\
 & GREG-R &  & 93.2 & 91.9 & 91.0 & 91.0 &  & 93.3 & 92.6 & 91.7 & 91.9 \\
CP & AB &  & 94.4 & 94.6 & 94.6 & 95.0 &  & 95.2 & 95.5 & 95.8 & 96.0 \\
 & ABL &  & 94.5 & 94.4 & 95.3 & 95.3 &  & 95.2 & 95.8 & 96.7 & 97.4 \\
 & ABH &  & 94.9 & 94.6 & 94.7 & 95.9 &  & 96.0 & 96.4 & 96.7 & 97.3 \\
 & HT &  & 95.9 & 95.9 & 95.9 & 95.9 &  & 95.5 & 95.5 & 95.5 & 95.5 \\
\hline
 & GREG &  & 7.01 & 6.87 & 6.73 & 6.58 &  & 7.27 & 7.12 & 6.96 & 6.79 \\
 & GREG-L &  & 7.12 & 7.09 & 7.07 & 7.05 &  & 7.40 & 7.37 & 7.35 & 7.32 \\
 & GREG-R &  & 7.09 & 6.98 & 6.88 & 6.81 &  & 7.36 & 7.24 & 7.14 & 7.05 \\
AL & AB &  & 7.47 & 7.57 & 7.67 & 7.78 &  & 7.80 & 7.92 & 8.06 & 8.24 \\
 & ABL &  & 7.49 & 7.60 & 7.72 & 7.86 &  & 7.80 & 7.93 & 8.06 & 8.22 \\
 & ABH &  & 7.45 & 7.54 & 7.65 & 7.78 &  & 7.75 & 7.83 & 7.94 & 8.06 \\
 & HT &  & 9.60 & 9.60 & 9.60 & 9.60 &  & 9.53 & 9.53 & 9.53 & 9.53 \\
\hline
\end{tabular}
\end{center}
\end{table}

\end{document}